\documentclass[%
 reprint,
 amsmath,amssymb,
 aps,
]{revtex4-2}

\usepackage{graphicx}
\usepackage{dcolumn}
\usepackage{bm}

\usepackage[version=4]{mhchem}
\usepackage{caption}
\usepackage{ragged2e}
\usepackage{subcaption}
\usepackage[dvipsnames]{xcolor}
\usepackage[normalem]{ulem}
\usepackage{comment}
\usepackage{titlesec}

\begin{document}

\preprint{APS/123-QED}

\title{Momentum imaging and kinetic energy release measurements for various fragmentation pathways in MeV energy proton collision with \ce{SO2} molecule}

\author{Sandeep Bajrangi Bari}
\affiliation{Department of Physics, Indian Institute of Technology Kanpur, Kanpur - 208016, India.}
\author{Ranojit Das}%
\affiliation{Department of Physics, Indian Institute of Technology Kanpur, Kanpur - 208016, India.}%
\author{R. Tyagi}%
\affiliation{Department of Physics, Indian Institute of Technology Kanpur, Kanpur - 208016, India.}%
\author{A. H. Kelkar}%
 \email{akelkar@iitk.ac.in}
\affiliation{Department of Physics, Indian Institute of Technology Kanpur, Kanpur - 208016, India.}%




\begin{abstract}

We have studied the ionization and fragmentation of \ce{SO2} molecular target in collision with 1 MeV proton beam using the technique of recoil ion momentum spectroscopy. Fragmentation dynamics of doubly charged \ce{SO2^{2+}} molecular ion has been investigated in detail using Dalitz plot and Newton diagrams. We have identified concerted and sequential dissociation pathways in three body dissociation of the parent molecular ion. 3D momentum distribution of all fragment particles, including the neutral atom, were obtained along with the kinetic energy release spectra for various fragmentation channels. 
 
\end{abstract}

\maketitle


\section{\label{sec:level1}Introduction\protect\\ }

Ionization and subsequent fragmentation of molecules in collisions with ions, electrons or photons is an area of fundamental as well as applied interest. Energy and charge transfer mechanism leading to various fragmentation channels govern the chemical evolution in environments with complex molecules. However, complete differential investigation of ionization and fragmentation dynamics of large molecules is a challenging task, theoretically and experimentally. Therefore, simple diatomic and triatomic molecules serve as model systems to study dissociation dynamics of molecular ions in gas phase. In comparison to diatomic molecules, the complexity of fragmentation dynamics grows multifold for a triatomic molecule. A triatomic molecular ion may undergo geometrical and symmetry changes accompanying two body and three body dissociation. There are countable number of triatomic molecule which exist in gas phase for experimentation. For example, linear triatomic molecules such as \ce{CO2} \cite{neumann2010fragmentation, lu2014evidence, khan2015observation, jana2012fragment, kushawaha2009polarization} and \ce{OCS} \cite{shen2016fragmentation, ramadhan2016ultrafast, kumar2018three} have been studied in detail however, only a few studies have focused on \ce{SO2}, which has bent geometry in the ground state \cite{cornaggia1996changes, hochlaf2004theoretical}.

\ce{SO2} plays an important role in the atmospheric chemistry of Earth and other planetary atmospheres. On Earth, it has both natural and manmade origins. In the presence of molecular oxygen and water, \ce{SO2} forms sulfuric acid in the upper atmosphere, contributing to harmful effects such as acid rains and smogs \cite{hu2013photochemistry}. In contrast, formation of sulfate aerosols due to \ce{SO2} emissions from industrial and volcanic activity may contribute toward atmospheric cooling by reflecting solar radiation back into the space thereby masking the greenhouse effect \cite{charlson1994sulfate, mitchell1997modification}. 
Sulfur dioxide is  also the most abundant gas, at 90 $\%$, in the atmosphere of Io, moon of planet Jupiter\cite{na1990international, barker1979detection, Lellouch2007}, which lacks \ce{CO2}. In the absence of carbon di oxide, \ce{SO2} may play a crucial role in the processes of \ce{O2} formation, thereby fixing the oxidation state of other elements in the atmosphere of Io. In fact, unlike \ce{CO2}, \ce{O2+} formation is a dominant channel in \ce{SO2} fragmentation \cite{kumar2024bond, duley2023fragmentation}.

Fragmentation dynamics of \ce{SO2^{q+}} molecular ions have been predominantly studied using photons \cite{Djuardin1981Photoion, curtis1985coincidence, eland1987dynamics, hsieh1997reaction, field1999fragmentation, hochlaf2004theoretical, jarraya2021state, wallner2022abiotic, masuoka2001kinetic,ben2005theoretical, masuoka2001single, lin2020femtosecond, salen2015complete}. Few groups have also investigated the dissociation dynamics with up to keV energy electrons \cite{basner1995electron, bhardwaj1999monte, masuoka1998dissociative, pal1998partial, chen2023fragmentation}. However, ion collision induced ionization and fragmentation of \ce{SO2} molecule has remained a relatively unexplored field \cite{rajput2020new}.

 In this work, we have studied collision of \ce{SO2} molecules in gas phase with 1 MeV proton beam and investigated the fragmentation dynamics of doubly and triply charged molecular ions. Two body and three body fragmentation pathways have been analyzed in detail using the techniques of Dalitz plot and Newton diagram. Unlike previous studies, we have also investigated the three body fragmentation channels involving neutral atoms and obtained kinetic energy release distribution and angular distribution for various dissociation pathways.

\begin{figure}[btp]
    \centering
    \vspace{-0.5cm}
    \includegraphics[width=\linewidth]{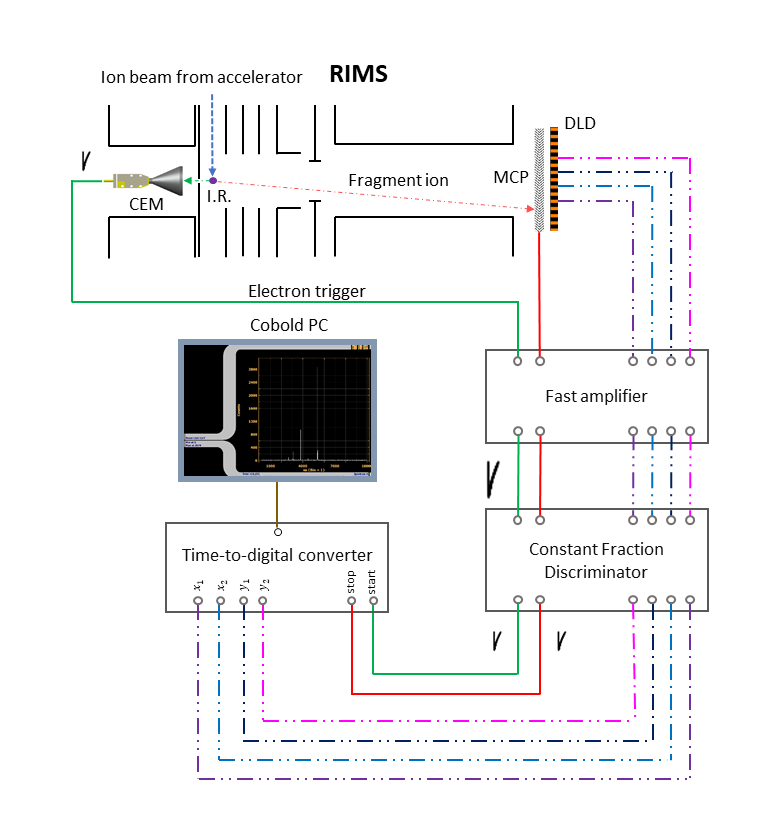}     
    \caption{ \justifying Schematic of the data acquisition system. The interaction region (I.R) is formed from the overlap of incident ion beam and effusive gas jet. The gas jet is normal to the figure.}
     \label{fig:DAS}
 \end{figure} 
 
\section{\label{sec:level2}Experimental details\protect}
The experiments were performed at the 1.7 MV Tandetron accelerator facility at IIT Kanpur. Fig.~\ref{fig:DAS}  shows the schematic of experimental set-up with data acquisition system. 1 MeV proton beam, obtained from the tandetron accelerator, was incident on an effusive jet of \ce{SO2} gas in a high vacuum scattering chamber. The ion beam - effusive gas jet interaction zone was situated symmetrically between the pusher and puller plates of a recoil ion momentum spectrometer (RIMS) developed in-house \cite{duley2022design}.

 The recoil ions and electrons generated from the collision were extracted, in opposite directions, by a uniform electric field normal to the extraction plane. The electron hit signal from a channel electron multiplier (CEM) at one end of the spectrometer acted as the trigger for the ToF measurement while the ion-signal from the MCP hit at the other end of the spectrometer served as the stop signal. A delay line anode placed behind the MCP was used to measure the ion hit position on the MCP. The CEM, MCP and delay line signals were amplified using a fast amplifier and fed to a constant fraction discriminator (CFD) for generating nim standard timing signals. The CFD outputs were further fed to a multi-hit time to digital converter (TDC) connected to a PC. The Cobold PC software (CoboldPC 2011 R5-2-x64 version 10.1.1412.2, Roentdek Handels GmbH, Frankfurt, Germany) was used to store time and position data for each coincidence event in a list mode file. An analysis code was written in the Cobold PC to generate fragment ion momenta and kinetic energy release distributions. The background pressure in the scattering chamber was maintained below $1.0\times10^{-7}$ mbar while the operating pressure was maintained at $\sim 1.1\times10^{-6}$ mbar throughout the experiment. The H$^+$ beam current was kept at $\sim$ 180 pA to ensure single collision condition and to keep the MCP count rate at $\sim$ 7 kHz. The low ion current and MCP count rate were maintained to avoid pile up effects which may result in MCP saturation. The data was taken in the list mode format (.lmf) for up to three ion hits on the MCP within a time window of 10 $\mu s$.

 The three-dimensional momenta of each particle were reconstructed from the obtained data using the following formulae :
 \begin{align}
     p_x &= \frac{m(x-x_0)}{fT_L}\label{eqn: px}\\
     p_y &= \frac{m(y-y_0)}{fT_L}\label{eqn: py}\\
     p_z &= -CqE_x\Delta T_L\label{eqn: pz}
      \end{align}
Here, $f$ and $C$ are magnification factors of the RIMS \cite{duley2022design}. $(x,y)$ are coordinates of ion hit position and $(x_0, y_0)$ are the coordinates of the centroid of the interaction zone projected on the MCP. $T_L$ is the ToF of the ions and $\Delta T = T - T_0$, where $T_0$ is the ToF of recoil ions with $p_z = 0$. 

The K.E. of a particle in an event is given by
\begin{align}
    K.E. = \sum_i\frac{p_x^2 + p_y^2 + p_z^2}{2m}
\end{align}
and the KER of an event is given by 
\begin{align}
    KER = \sum_n (K.E.)_n
\end{align}
where the sum is over all the particles in the event.

\section{\label{sec:level3}Results and discussion\protect}

\begin{table}
\begin{tabular}{c}
  \includegraphics[width=0.40\textwidth]{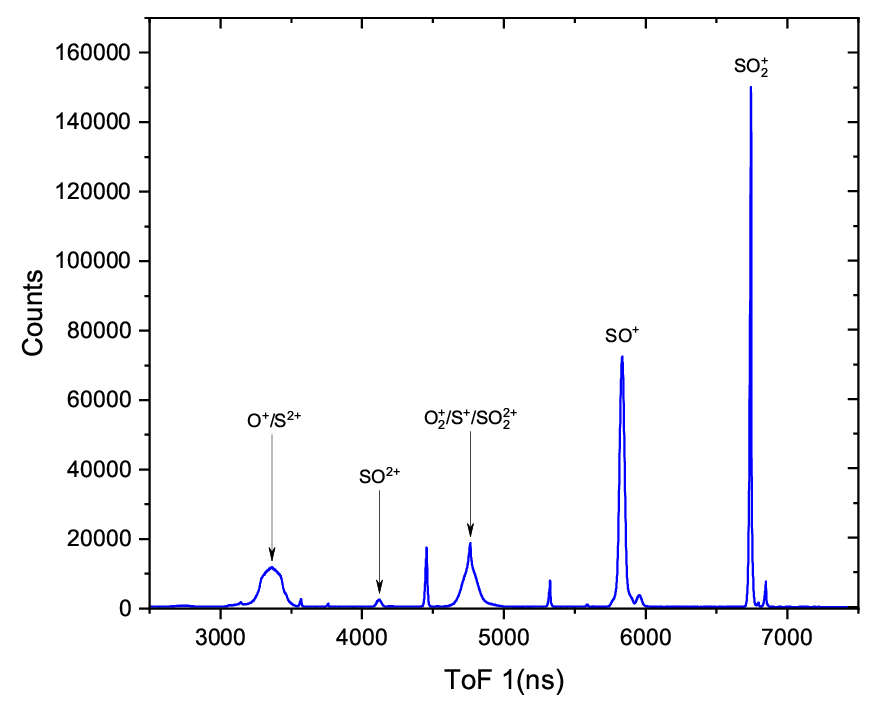}\\
  \hspace{1cm}(a)\\
      \\
  \hspace{1.3cm}\includegraphics[width=0.45\textwidth]{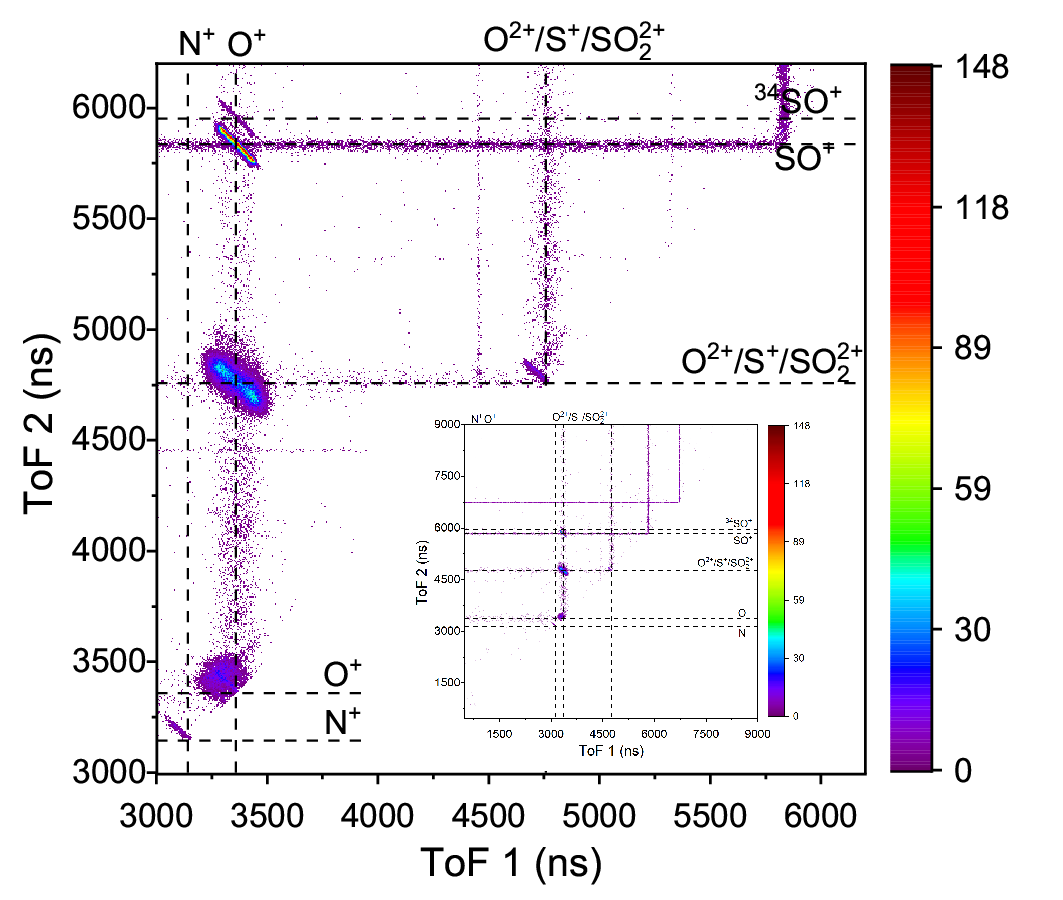}\\
  \hspace{1cm}(b)\\
\end{tabular}
     \captionof{figure}{ \justifying (a) Time of flight spectrum for the first ion obtained in the triple coincidence. Similar spectra were obtained for second ion hit and third ion hit. (b) Coincidence plot of ToF2 v/s ToF1. The inset shows the coincidence plot for the full range of ToF.}\label{fig: spec}
\end{table}

Fig.~\ref{fig: spec}(a) shows the time of flight (ToF) spectrum obtained for collision of 1 MeV proton with \ce{SO2} gas. In fig.~\ref{fig: spec}(b) we have shown the coincidence plot of ToF of second particle hit at the MCP vs ToF of first particle hit from the same collision event. Island like features in the plot correspond to various fragmentation channels. Islands corresponding to two-body break-up are easily identifiable in the coincidence plot with a characteristic slope $\sim$ -1, arising due to momentum conservation. However, for three body fragmentation channel where the central atom/ion receives zero momentum, the slope of the island in the coincidence plot is also $\sim$ -1. In such cases, detection of the third particle alleviates the ambiguity.

Diffused traces in the coincidence plot always correspond to three body fragmentation (for a tri-atomic molecular ion). Similar plots of ToF 3 versus ToF 2 and ToF 1 versus ToF 3 were also obtained. In general, diffused traces with no corresponding traces in the remaining two coincidence plots imply undetected third particle in the three body fragmentation channel. These events correspond to three body break-up involving one neutral particle.
 
\begin{table*}
\begin{tabular}{ccc}
  \includegraphics[width=0.32\textwidth]{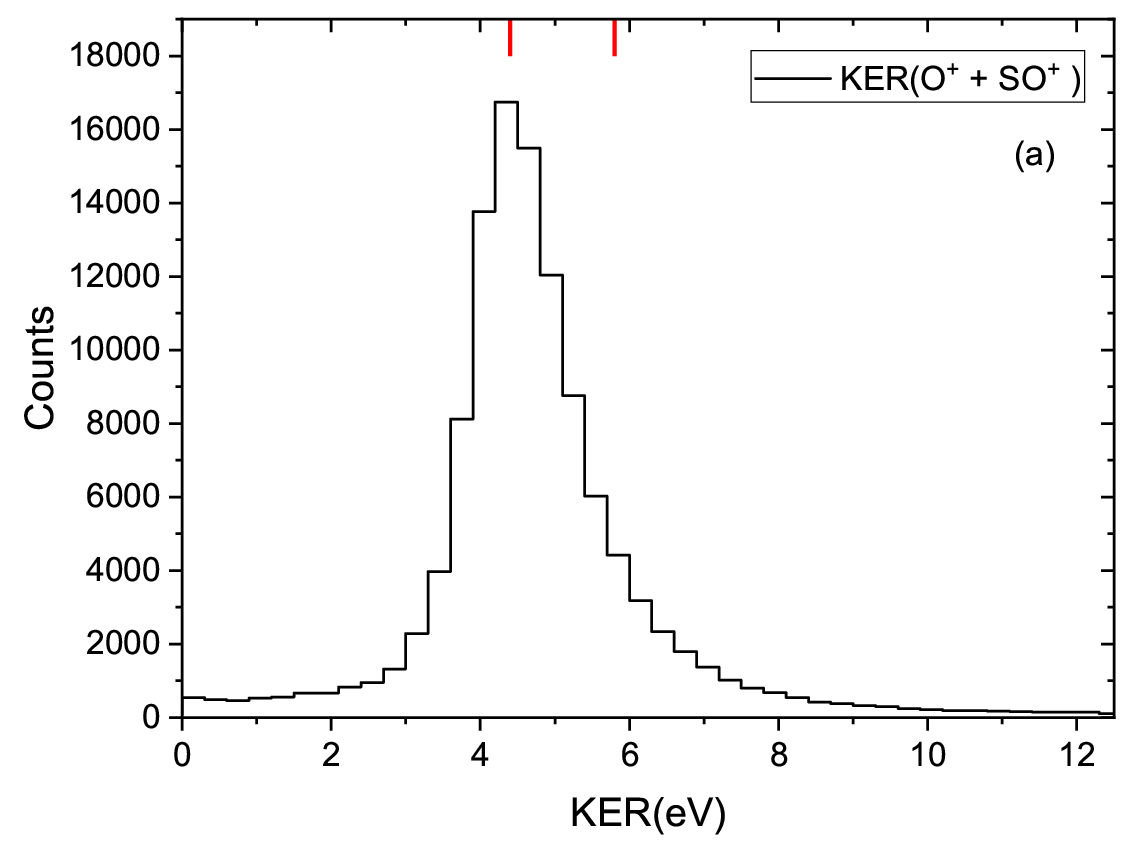} &
  \includegraphics[width=0.32\textwidth]{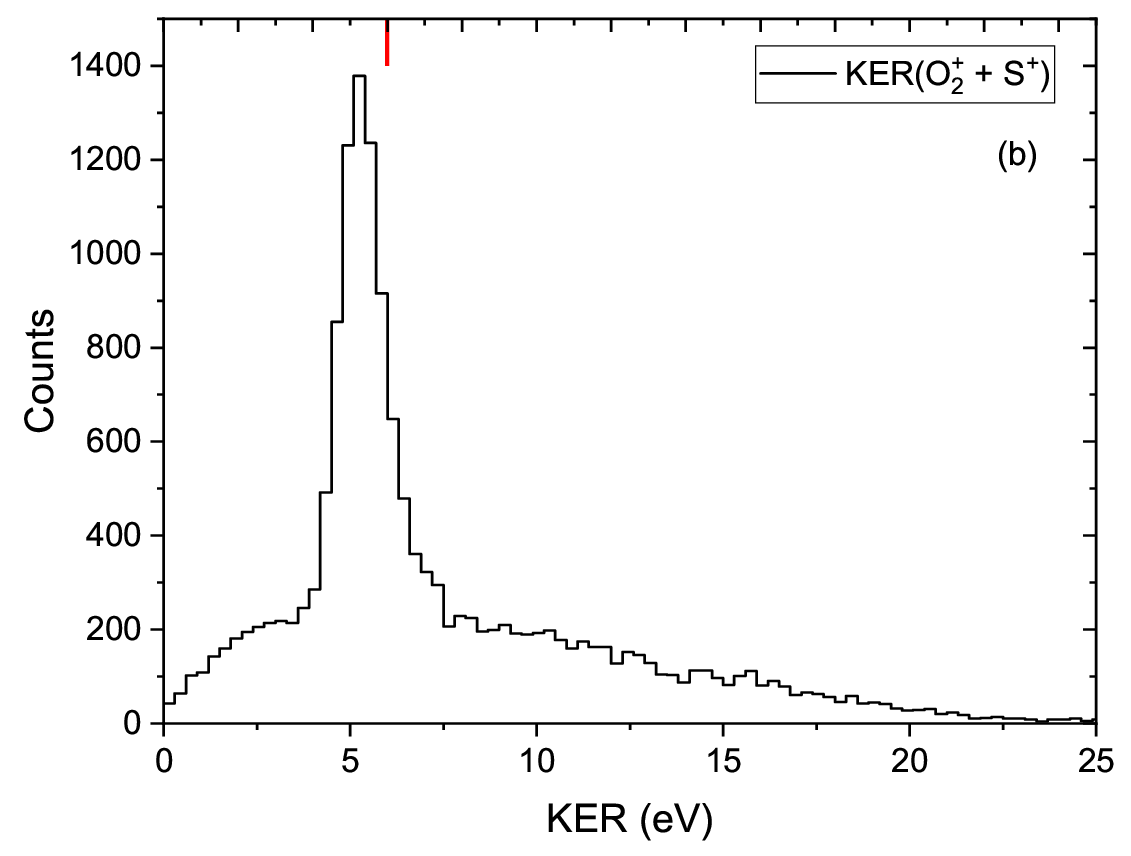} &
  \includegraphics[width=0.31\textwidth]{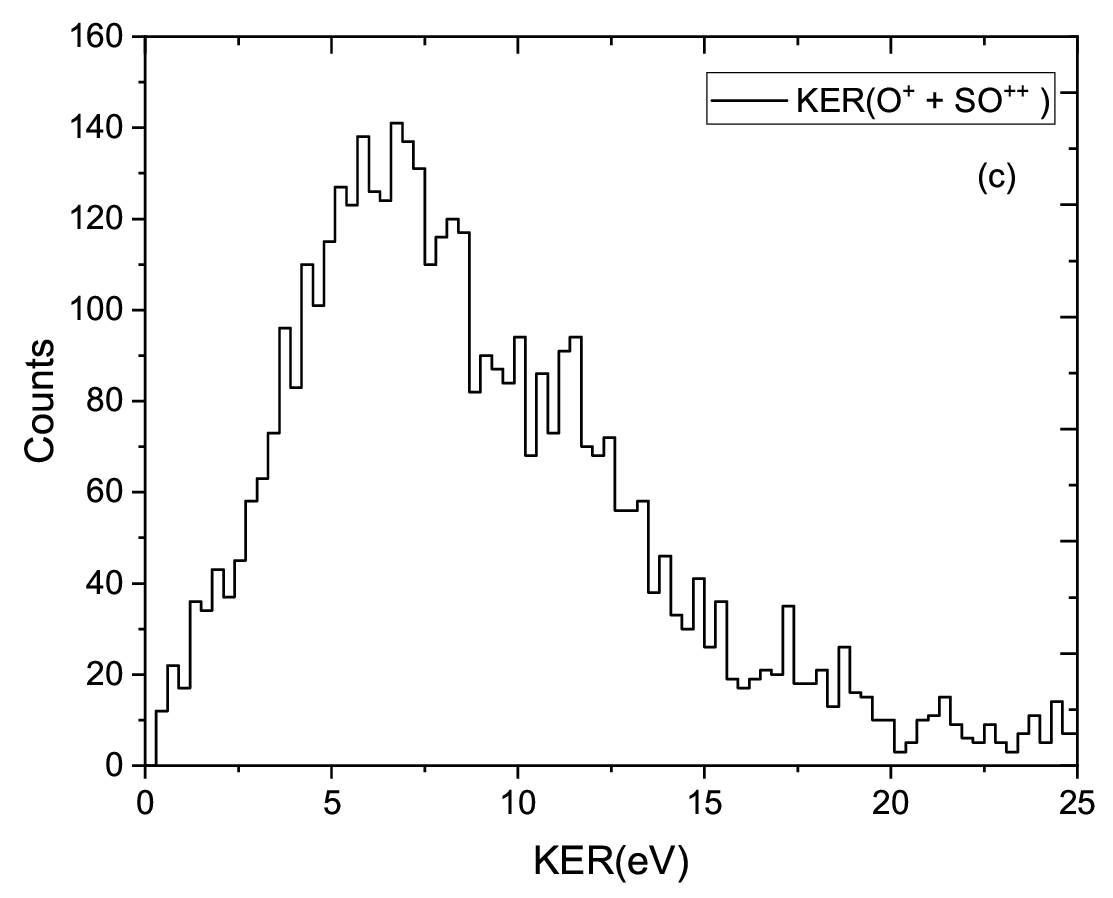} \\
\end{tabular}

     \captionof{figure}{ \justifying Kinetic energy release distribution for double fragmentation. (a) Channel \ce{ SO2^2+  -> O+ + SO+}. KER values of 4.4 eV and 5.8 eV have been designated in red markers. (b) Channel \ce{ SO2^2+  -> S+ + O2+}. The KER values of 6.0 eV is shown in red marker. (c) Channel \ce{ SO2^3+  -> O+ + SO++}}\label{fig: Double fragmentation}
\end{table*}

From the coincidence plot, fig.~\ref{fig: spec}, we have identified six fragmentation pathways. These can be separated as, two body fragmentation channels:
\begin{align*}
\ce{ SO2^2+  &-> O+ + SO+} \\
\ce{ SO2^2+  &-> S+ + O2+} \\
\ce{ SO2^3+  &-> O+ + SO++}
\end{align*}
three body fragmentation channels with one neutral:
\begin{align*}
\ce{SO2^{2+}  &-> O+ + O+ + S}\\
\ce{SO2^{2+}  &-> O+ + S+ + O}
\end{align*}
and three body fragmentation channel with all ionic fragments:
\begin{align*}
\ce{SO2^{3+}  -> O+ + O+ + S+}
\end{align*}
Neutral particles cannot be detected at the MCP and therefore momenta of the neutral particles for relevant collision events were deduced from those of the remaining two detected charged particles using conservation of momentum. 

Two body dissociation of \ce{SO2^{q+}, q = 2, 3} has been studied in detail earlier. In fig.~\ref{fig: Double fragmentation}(a, b and c) we have shown the KER distribution for the three fragmentation channels, (a) \ce{ SO2^2+  -> SO+ + O+}, (b) \ce{ SO2^2+  -> S+ + O2+}, and (c) \ce{ SO2^3+  -> SO^{2+} + O+} respectively. The KER distribution and the most probable KE values match well with those reported in literature \cite{curtis1985coincidence, eland1987dynamics, field1999fragmentation, masuoka2001kinetic, dujardin1984double, chen2023fragmentation, jarraya2021state}. For \ce{ SO2^2+  -> SO+ + O+} channel, the 4.5 eV peak is attributed to the decay of \ce{ SO2^{2+}}($^1A^\prime$) to \ce{SO2^{+}($X ^2\Pi $) + O+($^4 S^u $)} and the value of 5.1eV is attributed to the decay of $\ce{ SO2^{2+}}$($^1A^\prime$) to \ce{ SO2^{+}($X ^2\Pi $) + O+($^4 S^u $)} via crossings between $^1A^\prime$ and $^3A^\prime/^3A^{\prime\prime}$ states. The calculated theoretical values are  4.4 eV and 5.8 eV respectively. The fragmentation pathway \ce{ SO2^2+  -> S+ + O2+} involves isomerization of \ce{SO2^{2+}} resulting in the formation of \ce{O2+} \cite{wallner2022abiotic}. It has been shown that starting from a linear geometry, which is also the ground state geometry of \ce{[OSO]^{++}} \cite{hochlaf2004theoretical}, \ce{[OSO]^{++}} molecular ion converts to a bent geometry due to roaming mechanism of one of the oxygen atoms around the \ce{SO} bond and then again back to linear with both \ce{O} atoms on the same side of Sulfur atom forming \ce{[OOS]^{++}}. Further evolution of the system then leads to fragmentation into \ce{S+} and \ce{O2+}. For the channel \ce{ SO2^3+  -> SO^{2+} + O+}, the KER peaks at 6.6 eV and 11.4 eV.

\begin{table}
\begin{tabular}{c}
  \includegraphics[width=0.46\textwidth]{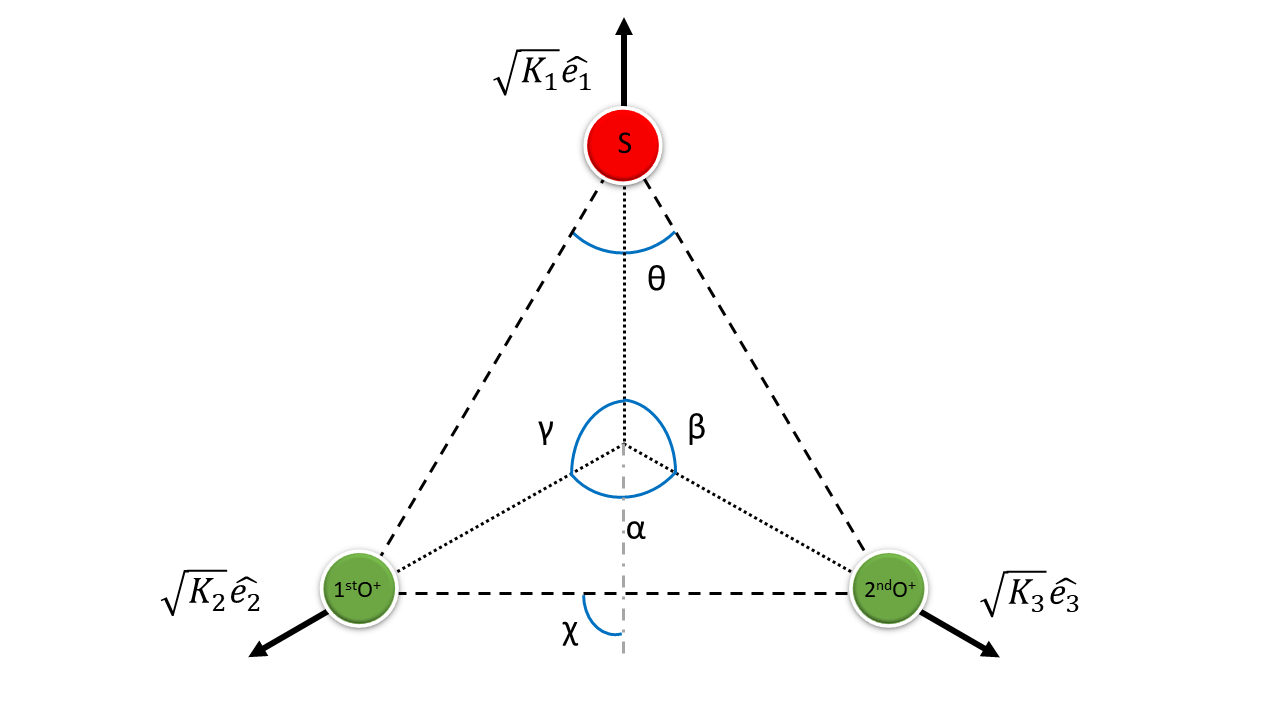}\\
  (a)\\
      \\
  \includegraphics[width=0.25\textwidth]{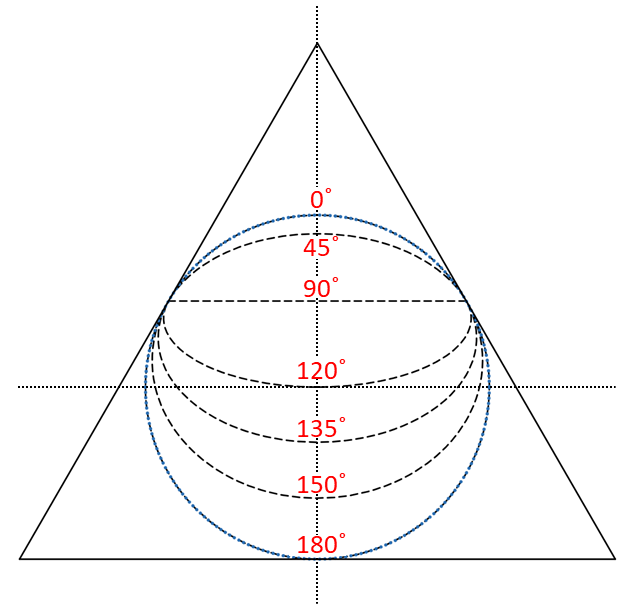}\\
  (b)\\
\end{tabular}
     \captionof{figure}{ \justifying (a) The different correlation angles between the momentum vectors at the instant of fragmentation. The bond angle is defined by angle $\alpha$. $\theta$ is the momentum space molecular bond angle (b) Curves of constant angles $\alpha$ between the vectors  $ \sqrt{K_2} \hat{e_2}$ and $ \sqrt{K_3} \hat{e_3}$}\label{fig: DMAC}
\end{table}

Unlike two body dissociation, the analysis of three body dissociation channels is more involved. Three body fragmentation dynamics can be analyzed using the Dalitz plot formalism, where reduced energies of the fragmented particles are plotted in a triangular coordinate system. The reduced energies of the fragmented particles in the center of mass frame coordinates are defined as
 \begin{equation}
    K_i = \frac{p{_i}^2}{\sum_{i}^{} p{_i}^2} 
 \end{equation}  
 so that
 \begin{equation}
  \sum_{i}^{} K{_i} = 1 
 \end{equation}
 The sum of the normals from any interior point, on the sides of an equilateral triangle of height unity also equals unity. Therefore, a point in the interior of the equilateral triangle is used to represent the coordinates $ K_i $, where the magnitude of normal from the interior point to the sides gives the values of $ K_i $. Projection on the Cartesian co-ordinate axes, with center of the triangle as origin, gives the Dalitz co-ordinates as
\begin{equation}
 K_x = \frac{K_3}{\sqrt{3}} - \frac{K_2}{\sqrt{3}}
 \end{equation}
 \begin{equation}
 K_y = K_1 - \frac{1}{3} 
 \end{equation}
while the inverse transformation equations are
\begin{equation}
 K_1 = K_y + \frac{1}{3}
\end{equation}
 \begin{equation}
 K_2 = - \frac{\sqrt{3}}{2}K_x - \frac{K_y}{2} + \frac{1}{3}
\end{equation}
 \begin{equation}
 K_3 =   \frac{\sqrt{3}}{2}K_x - \frac{K_y}{2} + \frac{1}{3}
\end{equation}
Conservation of momentum $\sum_{i}^{} \vec{p{_i}} = 0$ demands that all the points $(K_1, K_2, K_3)$ lie on or in the incircle of the equilateral triangle. Each point in a Dalitz plot gives a unique angular correlation between the momentum vectors in the center of mass frame of the molecule, thus representing the fragmentation geometry of the molecule at the instant of break-up. At any point in the Dalitz plot the angular correlation between the momentum vectors $\sqrt{K_2} \hat{e_2}$ and $\sqrt{K_3} \hat{e_3}$ is given by
 \begin{equation}
  \alpha = \arccos\left(\frac{2K_y - \frac{1}{3}}{2 \sqrt{(\frac{1}{3} - \frac{K_y}{2} - \frac{\sqrt{3}}{2}K_x) (\frac{1}{3} - \frac{K_y}{2} + \frac{\sqrt{3}}{2}K_x)}}\right) \label{eqn:alpha}
 \end{equation}
 where $\hat{e_i} = \vec{p_i}/|\vec{p_i}|$ while those between $\sqrt{K_1} \hat{e_1}$ and $\sqrt{K_3} \hat{e_3}$ \& $\sqrt{K_1} \hat{e_1}$ and $\sqrt{K_2} \hat{e_2}$ are given by
 \begin{equation}
 \beta = \arccos\left(\frac{- \sqrt{3}K_x - K_y - \frac{1}{3}}{2 \sqrt{(\frac{1}{3} + K_y )(\frac{1}{3} - \frac{K_y}{2} + \frac{\sqrt{3}}{2}K_x)}}\right) \label{eqn:beta}
 \end{equation}
 \begin{equation}
 \gamma = \arccos\left(\frac{\sqrt{3}K_x - K_y - \frac{1}{3}}{2 \sqrt{(\frac{1}{3} + K_y )(\frac{1}{3} - \frac{K_y}{2} - \frac{\sqrt{3}}{2}K_x)}}\right) \label{eqn:gamma}
\end{equation}
 In fig.~\ref{fig: DMAC}(a), we define the correlation angles between the momentum vectors of the fragmentation products. Eqns. (\ref{eqn:alpha}) - (\ref{eqn:gamma}) show that fixed angular correlations between two fragmentation products lie on a curve in the Dalitz plot.
 Fig.~\ref{fig: DMAC}(b) shows the curves of constant angles between the vectors $ \hat{e_2}$ and $\hat{e_3}$ in the phase space. 
 It should be noted that conservation of momentum renders one of the vectors $ \sqrt{K_i} \hat{e_i}$ redundant and only two quantities are needed in Dalitz plot. This allows analysis of fragmentation channels where one of the three fragmentation products is neutral which is not detected. The momentum of the neutral particle can be determined using conservation of total momentum. In the following sections, we shall discuss the three body fragmentation channels of \ce{SO2^{3+}} and \ce{SO2^{2+}} molecular ion. The fragmentation of \ce{SO2^{3+}} molecular ion results in three ionic fragments whereas the three body fragmentation channel associated with \ce{SO2^{2+}} molecular ion includes one neutral species.

 \subsection{\label{sec:level4}Three body fragmentation of \ce{SO2^{2+}}}
 \subsubsection{\label{sec:level5}\ce{SO2^{2+}  -> O+ + O+ + S}}

 \begin{figure}[btp]
    \centering
    \includegraphics[width=\linewidth]{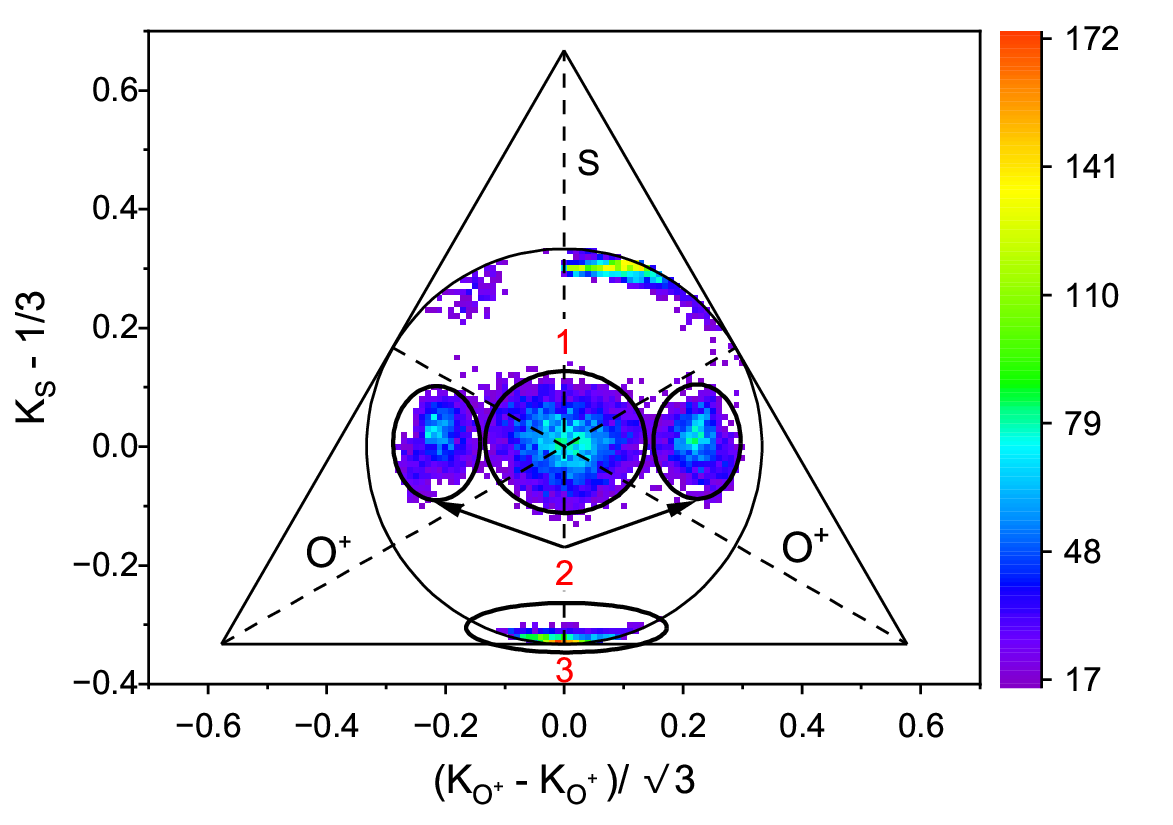}
    \caption{ \justifying Dalitz plot for triple fragmentation of \ce{SO2^2+ -> O+ + O+ + S}. Different regions correspond to different mechanisms of fragmentation.The regions have been marked for convenience.} 
     \label{fig:O+O+S_DP}
 \end{figure}
 
 Fragmentation of doubly charged \ce{SO_2^{2+}} molecular ion into two charged oxygen ions and a neutral sulphur atom is the dominant three body dissociation channel. Fig.~\ref{fig:O+O+S_DP} shows the Dalitz plot for this channel.

 The momentum of the neutral Sulhpur atom is calculated using conservation of momentum
 \begin{equation}
    \vec{p}_{\ce{S}} = - \left( \vec{p}_{\ce{O+}}  + \vec{p}_{\ce{O+}} \right) 
 \end{equation}
In the ground state, neutral \ce{SO2} molecule has bent geometry with angle of 119.5$^{\circ}$ between the two \ce{SO} bonds \cite{dujardin1984double}. The central region of the Dalitz plot (region 1, fig \ref{fig:O+O+S_DP}) represents fragmentation events with equal momentum sharing between the fragmenting ions and neutral atom.  The angular correlation is also preserved in this break-up channel. Therefore the fragmentation follows neutral molecular geometry. Region 2 in fig.~\ref{fig:O+O+S_DP} corresponds to asymmetric momentum sharing between the two \ce{O+} ions. However, the events 
are concentrated in the central part of the Dalitz plot implying that the bond angle is that of the neutral molecule. In region 3, fig. \ref{fig:O+O+S_DP}, the momentum of neutral sulfur atom is close to zero. This corresponds to fragmentation in a linear geometry. For \ce{SO2} molecular ion the linear geometry is attributed to the removal of electrons from 8a$_1$ orbital \cite{cornaggia1996changes}.
The different regions in the Dalitz plot (fig.~\ref{fig:O+O+S_DP}) will be discussed in detail in the following sections.

\begin{table*}
\begin{tabular}{ccc}
  \raisebox{0.07\height}{\includegraphics[width=0.283\textwidth]{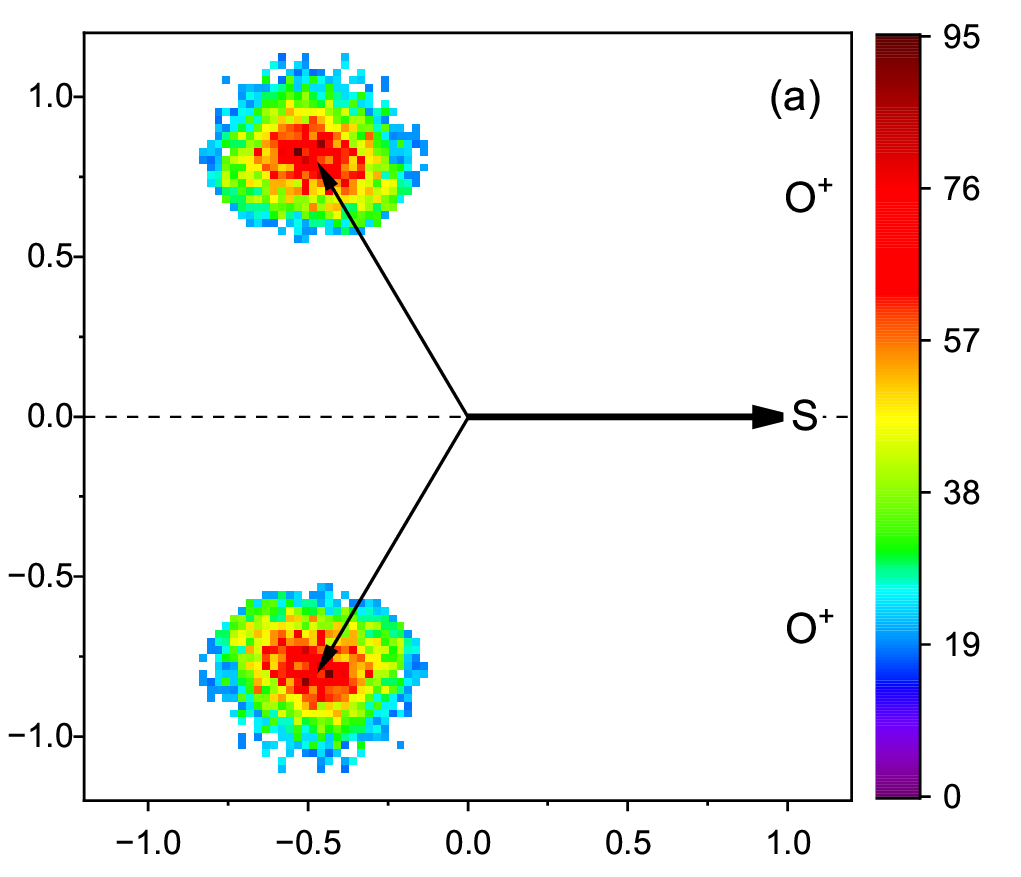}}&
  \includegraphics[width=0.33\textwidth]{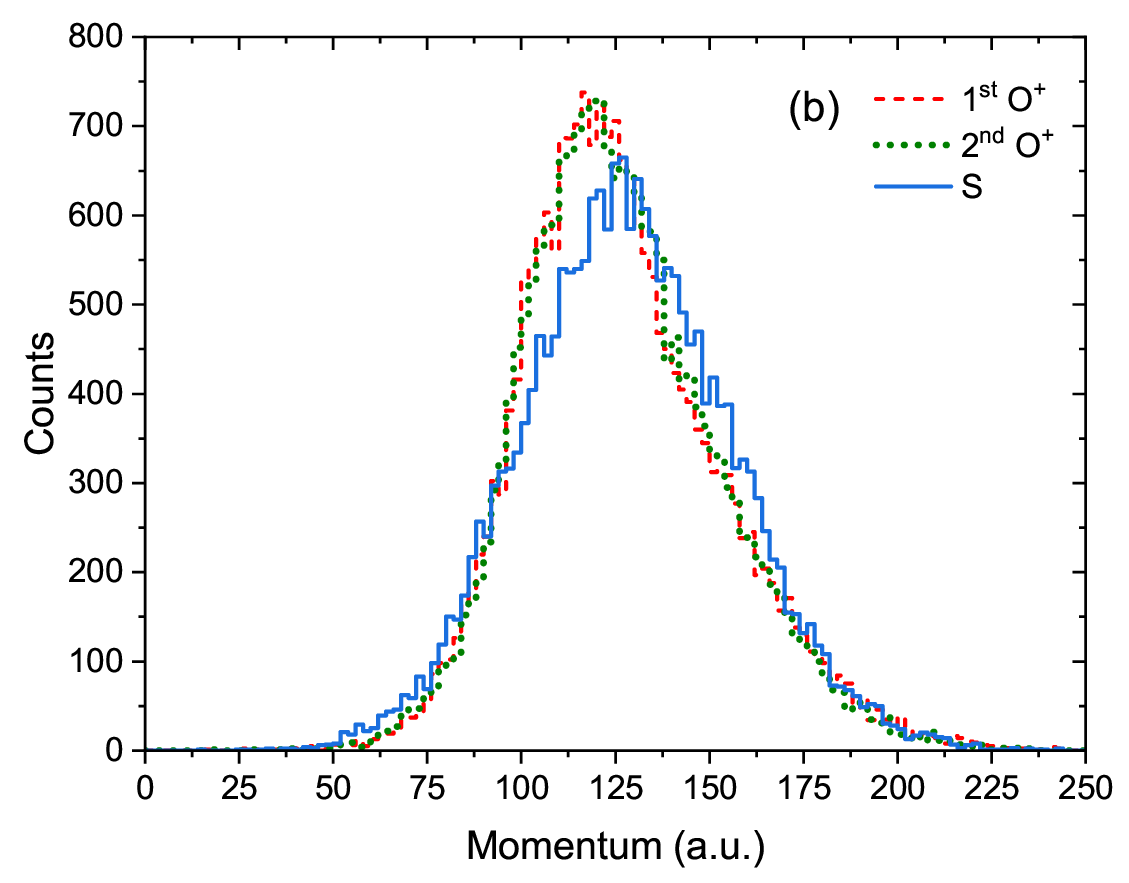} &
  \includegraphics[width=0.33\textwidth]{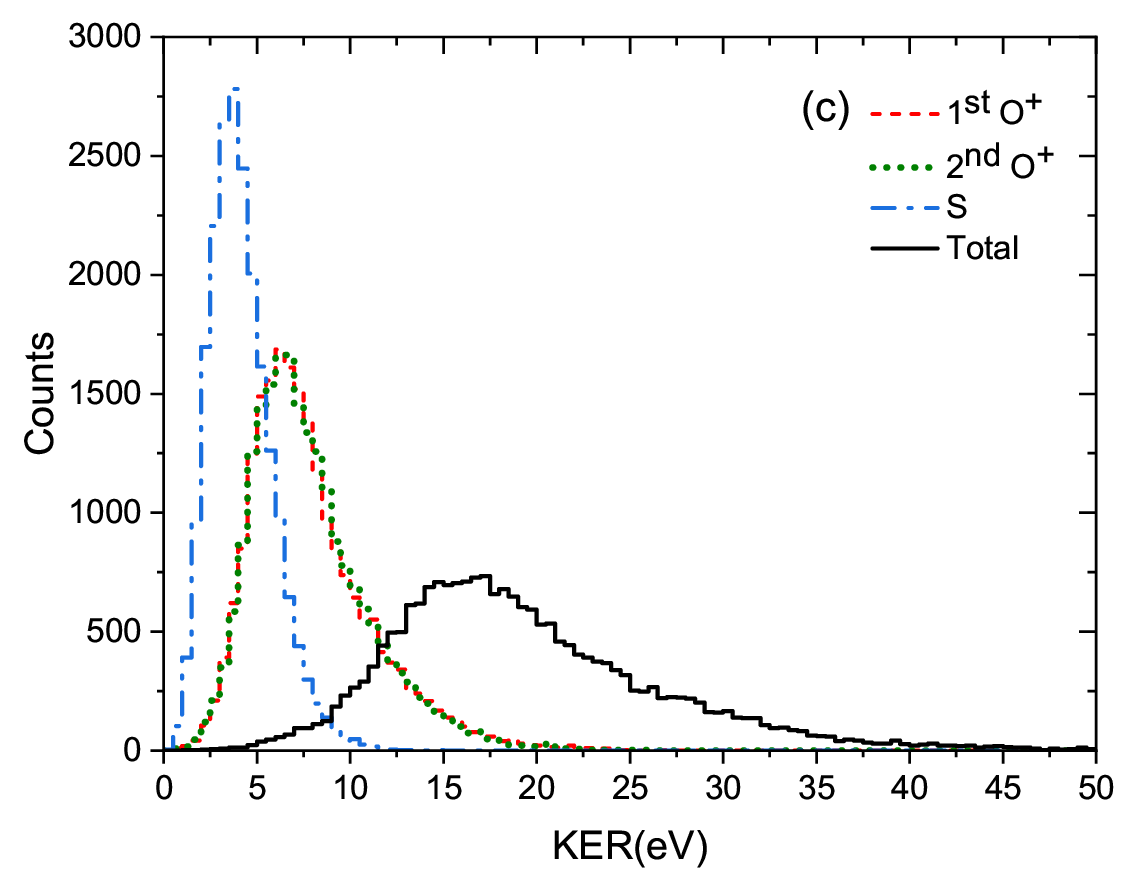} \\
\end{tabular}
     \captionof{figure}{ \justifying (a) Newton diagram for Region 1 of channel \ce{SO2^{2+}  -> O+ + O+ + S}, (b) Momentum distribution for the channel \ce{SO2^{2+}  -> O+ + O+ + S} and (c) KER distribution for the channel \ce{SO2^{2+}  -> O+ + O+ + S}}\label{fig: O+O+S_R1}
\end{table*}
\begin{table*}[ht]
\begin{tabular}{ccc}
  \raisebox{0.07\height}{\includegraphics[width=0.29\textwidth]{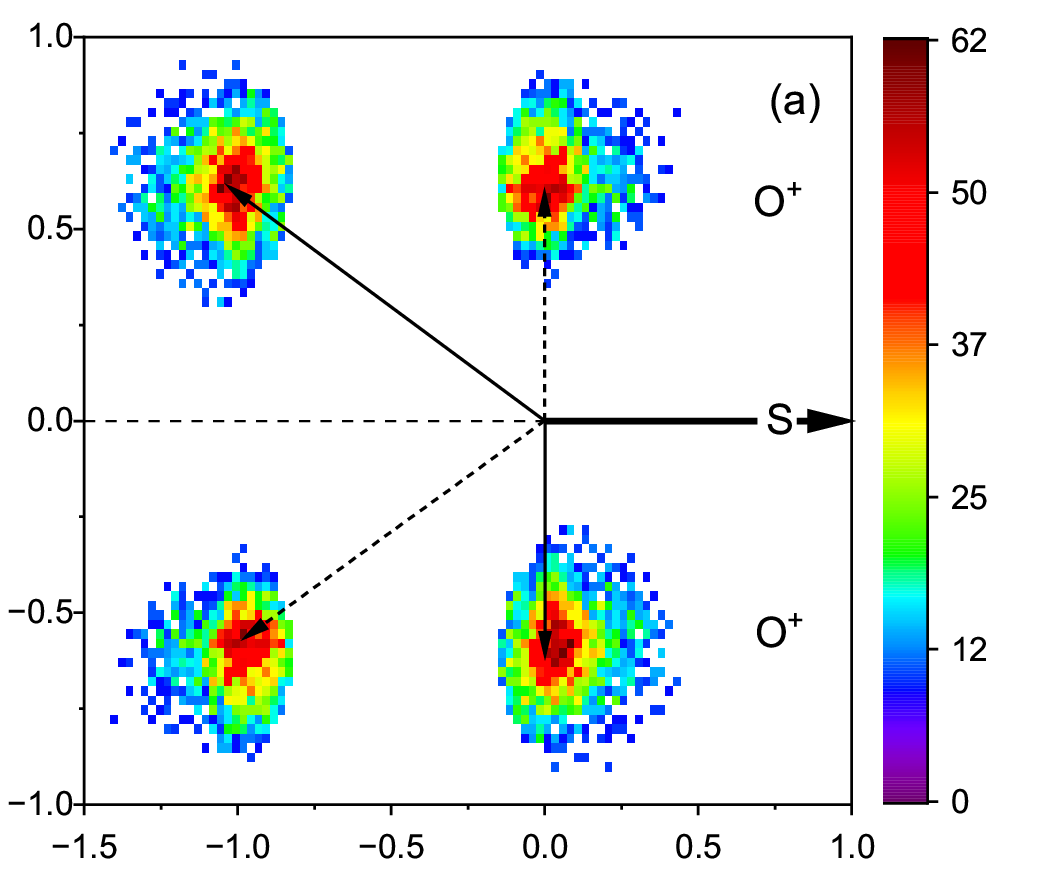}}&
  \includegraphics[width=0.33\textwidth]{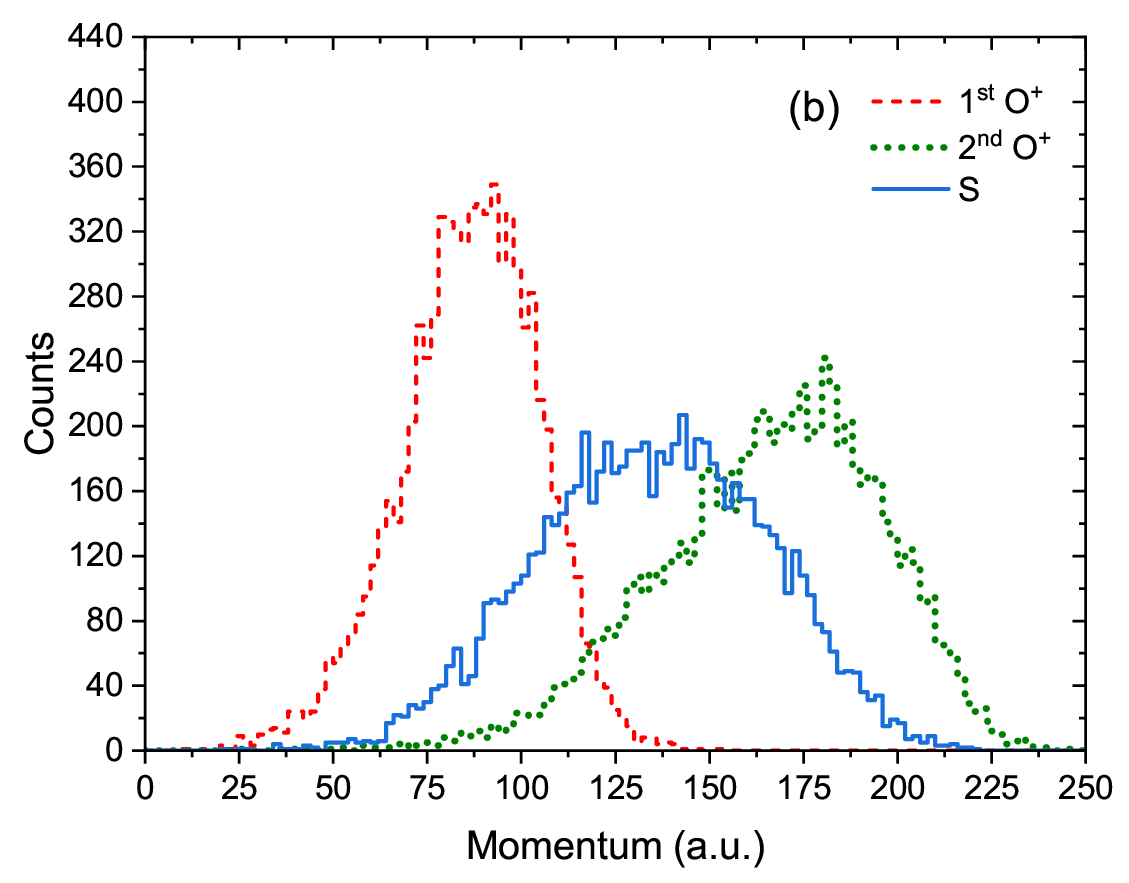} &
  \includegraphics[width=0.33\textwidth]{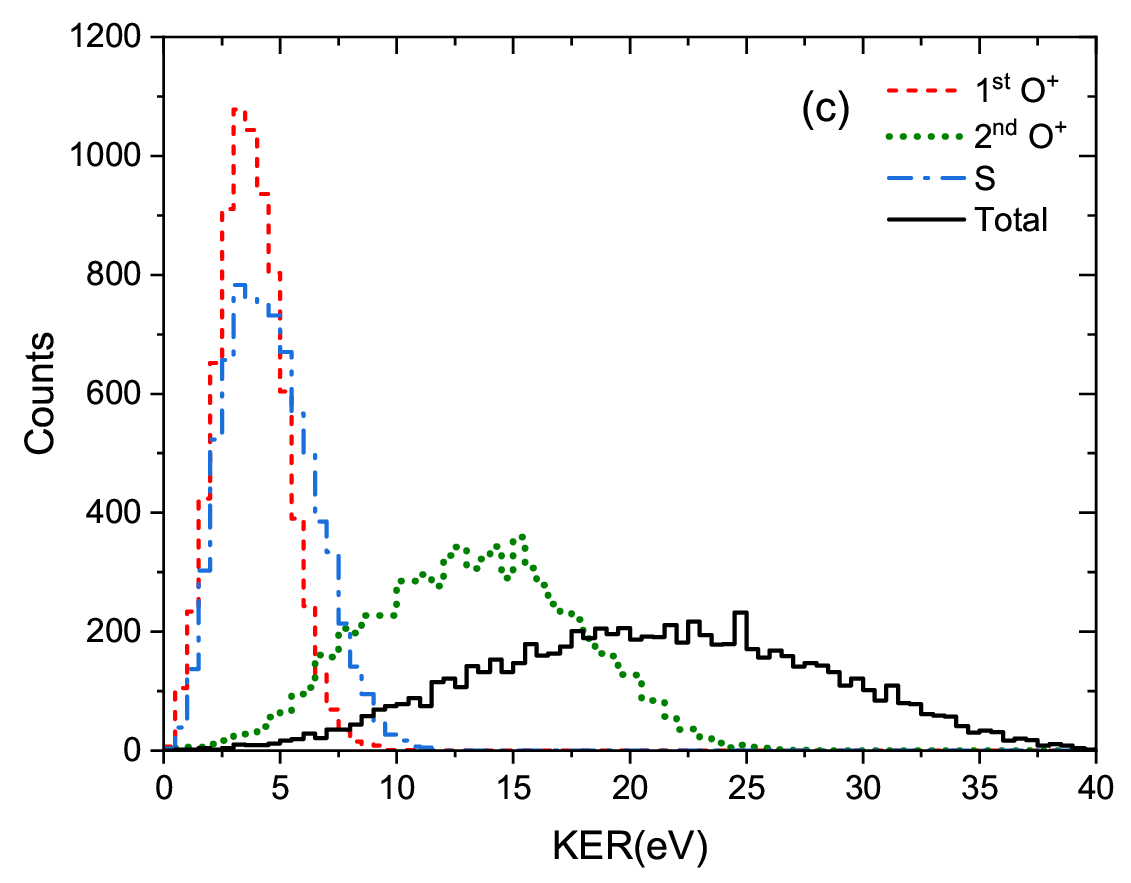} \\
\end{tabular}
     \captionof{figure}{ \justifying (a) Newton diagram for region 2 of channel \ce{SO2^{2+}  -> O+ + O+ + S}. The diagram contains event distributions from the lobes of region 2. (b) Momentum distribution for left lobe of Region 2 of the channel \ce{SO2^{2+}  -> O+ + O+ + S} and (c) KER distribution for left lobe of Region 2 the channel \ce{SO2^{2+}  -> O+ + O+ + S}}\label{fig: O+O+S_R2}
\end{table*}

\begin{center}
    \it{Region 1}
\end{center}

In region 1, the distribution is centered at the origin of the Dalitz plot. This is a characteristic feature of concerted fragmentation. In concerted fragmentation, all the bonds of the molecular ion break simultaneously. In figs.~\ref{fig: O+O+S_R1}(b) and ~\ref{fig: O+O+S_R1}(c), we show the total and individual momentum and KER distributions of the fragment ions and the neutral S atom. The fragment \ce{O+} ions and the neutral Sulfur atom carry equal momenta. Therefore, the distributions of all the momentum correlation angles peak at 120$^{\circ}$, which is approximately equal to the bond angle, 119.5$^{\circ}$ of the neutral \ce{SO2} molecule \cite{Djuardin1981Photoion}. This confirms that the geometry of the molecule is preserved during fragmentation. Therefore, the transition from neutral \ce{SO2} to \ce{SO2^{2+}} molecular ion occurs in the Franck-Condon region. However, this geometry is not the equilibrium geometry of the \ce{SO2^{2+}} ion which is linear in its ground state\cite{hochlaf2004theoretical}. The angular distribution of $\chi$ peaks at 90$^{\circ}$ as expected from the symmetric geometry and equal momentum sharing of both the \ce{O+} ions. The concerted nature of fragmentation is further verified by the Newton diagram for the region as shown in fig. \ref{fig: O+O+S_R1}(a).
\begin{table*}[htb]
\caption{\justifying The most probable values of the angular distribution of momentum correlation angles $\alpha, \beta, \gamma$, momentum space molecular bond angle $\theta$, angle $\chi$  and KER of the particles for the channel \ce{SO2^{2+}  -> O+ + O+ + S} (see fig.~\ref{fig: DMAC}(a) for more information).}
\label{Table: Angular distribution and KER O+O+S}
\begin{ruledtabular}
\begin{tabular}{cccccccccc}
\textrm{Region}\ &\textrm{$\alpha$} &\textrm{$\beta$} &\textrm{$\gamma$} &\textrm{$\theta$} &\textrm{$\chi$} &\textrm{KER(S)} &\textrm{KER(1$^{st}$O$^+$)} &\textrm{KER(2$^{nd}$O$^+$} &\textrm{Total KER}\\
\textrm{}\ &\textrm{degree} &\textrm{degree} &\textrm{degree} &\textrm{degree} &\textrm{degree} &\textrm{eV} &\textrm{eV} &\textrm{eV} &\textrm{eV}\\
\colrule
 1 & 120 & 120 & 120 & 60 & 90 & 3.5 & 6 & 6.5 & 17 \\
 2 (right) & 125 & 87 & 152 & 51 & 49 & 3.5 & 14.5 & 4 & 21 \\
 2 (left) & 123 & 148 & 90 & 50 & 130 & 3 & 3.5 & 15 & 21.5 \\
 3 & 173 & 109 & 97 & 167 & 106 &   & 3.5 & 3.5 & 7.5 \\
 \end{tabular}
 \end{ruledtabular}
 \end{table*}
From fig.~\ref{fig: O+O+S_R1}(c), we observe that both the oxygen ions have same kinetic energy distribution. Therefore, this is a case of synchronous concerted decay where the bonds break at the same rate in addition to breaking at the same instant \cite{strauss1990correlations}.

\begin{center}
   \it{Region 2}
\end{center}

  Region 2 consists of two clusters of events on either sides of region 1. The symmetry in Dalitz plot is also clearly reflected in the distribution of angle $\chi$ for the two regions, one peaking at 49$^{\circ}$ for the right lobe and the other at 130$^{\circ}$ for the left lobe. Figs.~\ref{fig: O+O+S_R2}(a, b and c) show the Newton diagram, momentum distribution and KER distribution for the left lobe of region 2. In fig.~\ref{fig: O+O+S_R2}(a), we also show the Newton diagram for the right lobe, however, due to the identical nature of momentum and KER distribution, these have not been shown for the right lobe. For both the lobes, the two \ce{O+} ions have different KER distributions, with one peaking at lower energy and the other peaking at a higher energy. The asymmetry in the KER distribution of the two \ce{O+} ions is reflective of asymmetric stretch of the two \ce{SO+} bonds. The symmetry-breaking in the KER distributions allows us to identify the two \ce{O+} ions as one with larger momentum and the other with lower momentum. However, the distributions of events for the two \ce{O+} ions are clustered in the Newton diagram so that even if the bonds do not break at the same instant, there is negligible rotation of the intermediate \ce{SO+} ions. Therefore, this is a case of asynchronous concerted decay \cite{strauss1990correlations}.

\begin{center}
    \it{Region 3}
\end{center}

\begin{table*}
\begin{tabular}{ccc}
  \raisebox{0.07\height}{\includegraphics[width=0.295\textwidth]{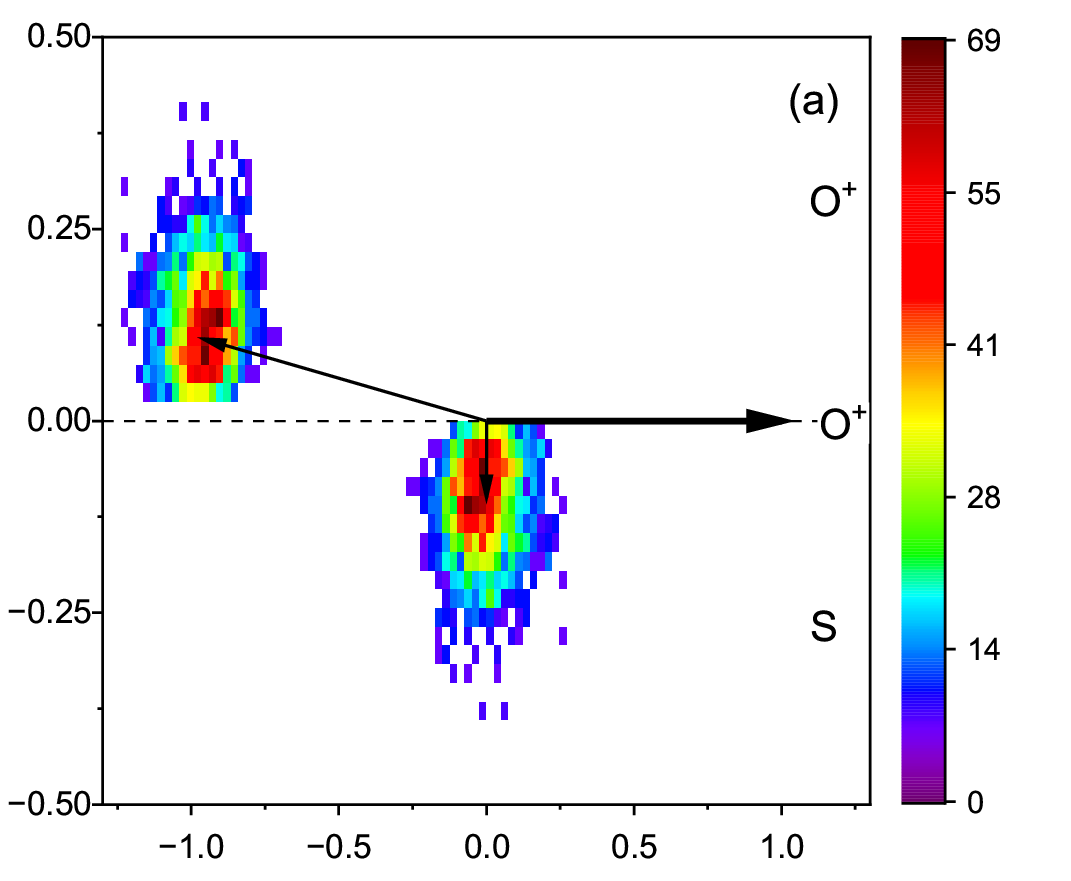}}&
  \includegraphics[width=0.325\textwidth]{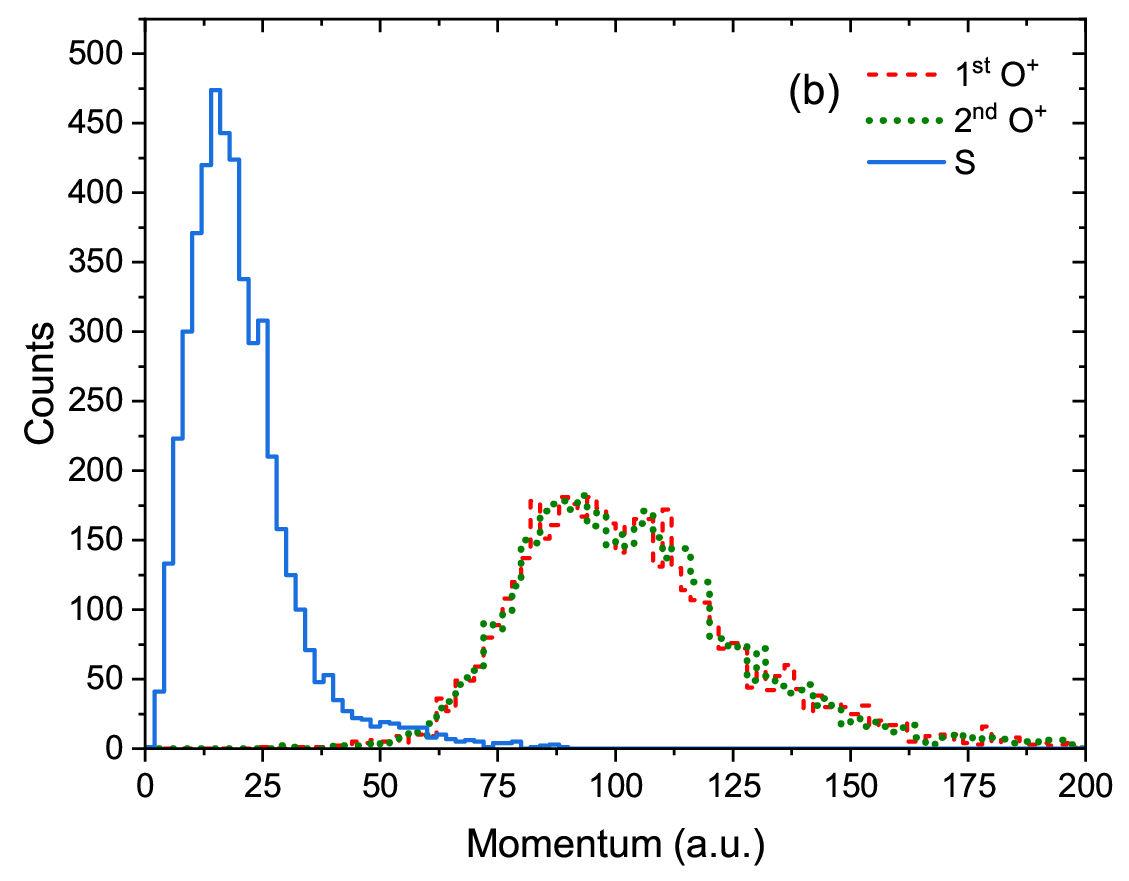} &
  \includegraphics[width=0.33\textwidth]{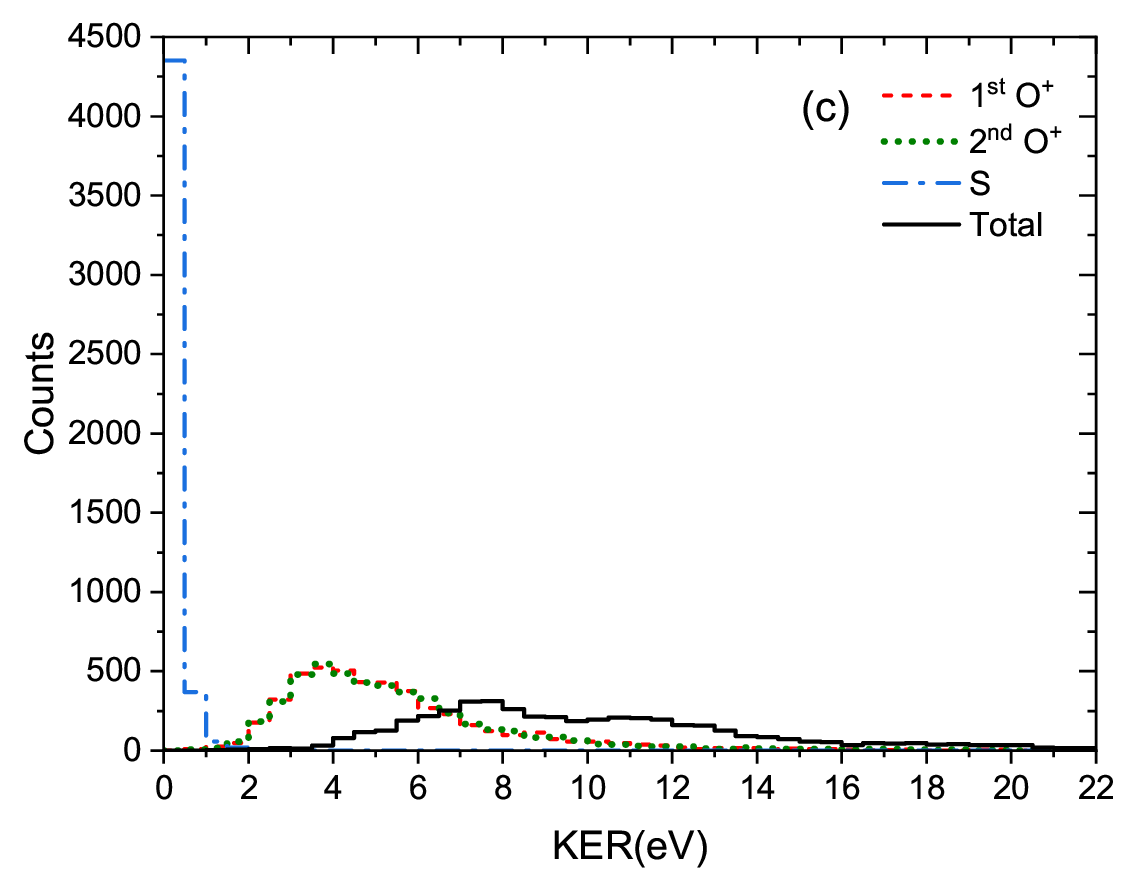} \\
\end{tabular}
     \captionof{figure}{ \justifying (a) Newton diagram for Region 3 of channel \ce{SO2^{2+}  -> O+ + O+ + S}, (b) Momentum distribution of Region 3 for the channel \ce{SO2^{2+}  -> O+ + O+ + S} and (c) KER distribution of Region 3 for the channel \ce{SO2^{2+}  -> O+ + O+ + S}}\label{fig: O+O+S_R5}
\end{table*}
Region 3 shows concentration of events at the base of the Sulfur axis. Events located in this region of Dalitz plot correspond to linear geometry (cf. fig. \ref{fig: DMAC}(b)) which is also the ground state geometry of \ce{SO2^{2+}} ion \cite{hochlaf2004theoretical}. Since the events accumulate on the Sulfur axis, therefore, this is a case of synchronous concerted breakup \cite{strauss1990correlations} where the two \ce{SO+} bonds break at the same instant and at the same rate.
The concerted nature of the fragmentation process is evident from the Newton diagram (see fig. \ref{fig: O+O+S_R5}(a)), where the distribution is rather concentrated. The linearity of the molecule is also established by back-to-back emission of \ce{O+} ions as reflected in Newton diagram and the momentum distribution of the two \ce{O+} ions (see fig. \ref{fig: O+O+S_R5}(b)). The \ce{O+} ions carry all the energy in the fragmentation process, thereby leaving the neutral S atom with nearly zero kinetic energy (see fig. \ref{fig: O+O+S_R5}(c)).\\ 
The strong momentum anti-correlation between the two \ce{O+} ions in region 3 suggests a strong correlation in the ToF coincidence plots of the two ions. Events of region 3 were directly gated out from the Dalitz plot and a coincidence plot of ToF of 2$^{nd}$ \ce{O+} ion v/s ToF of 1$^{st}$ \ce{O+} ion was constructed for region 3. The plot gives a slope of -0.91 which is close to value of -1.0 as expected from the momentum conservation in a two-body fragmentation. Therefore, momentum correlation between the terminal \ce{O+} ions in the three-body fragmentation of \ce{SO2^{2+}} molecular ion for region 3 is akin to that for two-body fragmentation. This, further establishes the linear geometry of the molecular ion in region 3.

\subsubsection{\label{sec:level6}\ce{SO2^{2+}  -> O+ + S+ + O}}
\begin{figure}[btp]
    \centering
    \includegraphics[width=\linewidth]{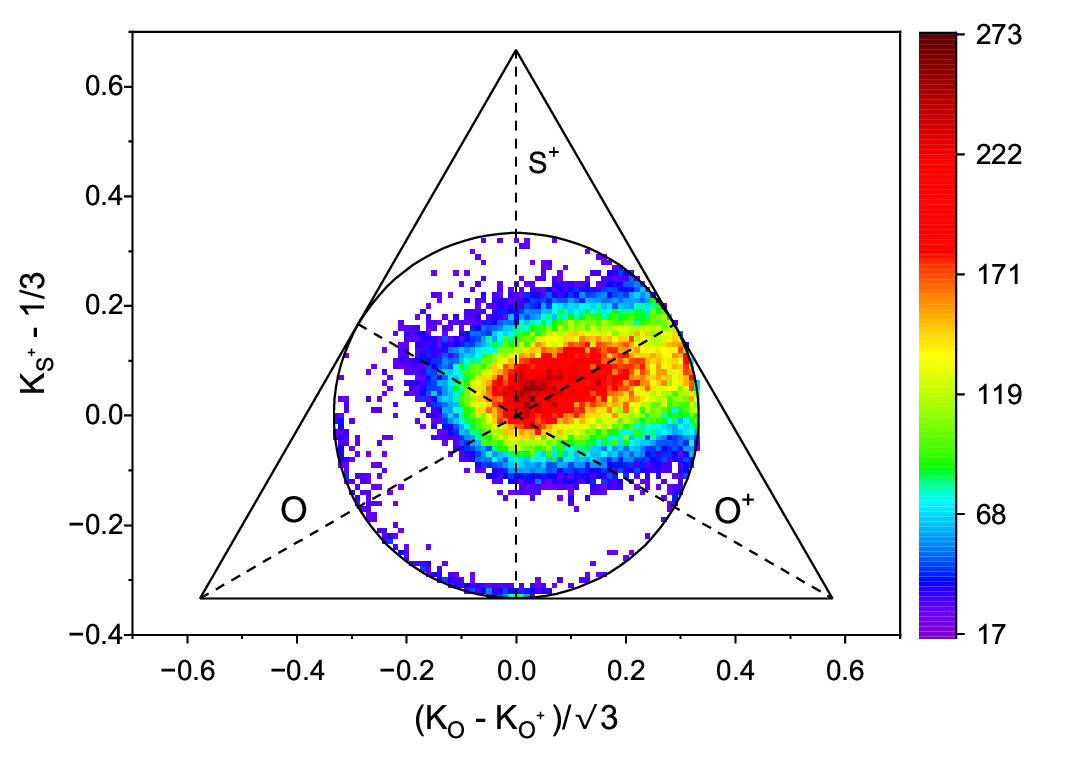}
    \caption{ \justifying Dalitz plot for the fragmentation channel \ce{SO2^{2+}  -> O+ + S+ + O}. The distribution of events results from sequential and concerted fragmentation mechanisms.} 
    \label{fig:O+S+O_DP}
\end{figure}

Fragmentation of \ce{SO2^{2+}} into \ce{O+} and \ce{S+} ions with a neutral oxygen atom \ce{O} is also a dominant dissociation pathway. In
fig. \ref{fig:O+S+O_DP}
we have shown the corresponding Dalitz plot. The distribution of events extends from the base of the neutral oxygen axis (right side of the Dalitz triangle) to the central region. This distribution of events corresponds to the bent geometry of the fragmenting molecular ion. The bond angle is centered at $\sim$ 120$^o$ (see \ref{fig: DMAC}(b)). This is also the ground state geometry of the neutral \ce{SO2} molecule \cite{dujardin1984double}. The fragmentation channel may have contributions from concerted as well as sequential dissociation pathways. The presence of a sequential dissociation pathway can be ascertained using the native-frame analysis \cite{rajput2018native}. In this analysis, we calculate the angle $\phi$ between $\vec{p}_{O^+S^+}$ and $\vec{p}_O$ momentum vectors for each event. We further plot a 2D correlation map of angle $\phi$ and the total kinetic energy of \ce{S+} and \ce{O+} fragment ions. The kinetic energy is calculated in the center of mass frame of the \ce{SO^{2+}} molecular ion. The native-frame plot is shown in fig. \ref{fig: O+S+O_R2}(a).
\begin{table*}
\begin{tabular}{ccc}
  \includegraphics[width=0.349\textwidth]{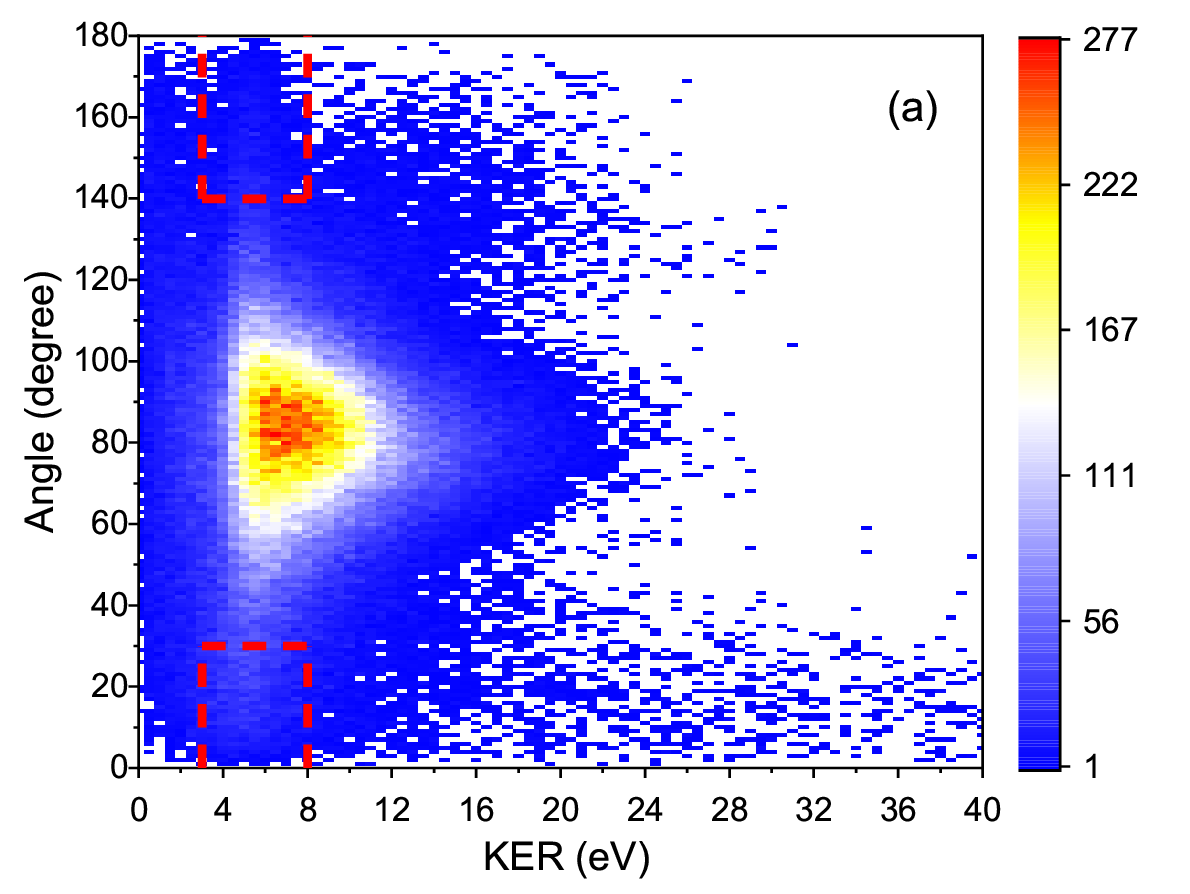} &
 \includegraphics[width=0.315\textwidth]{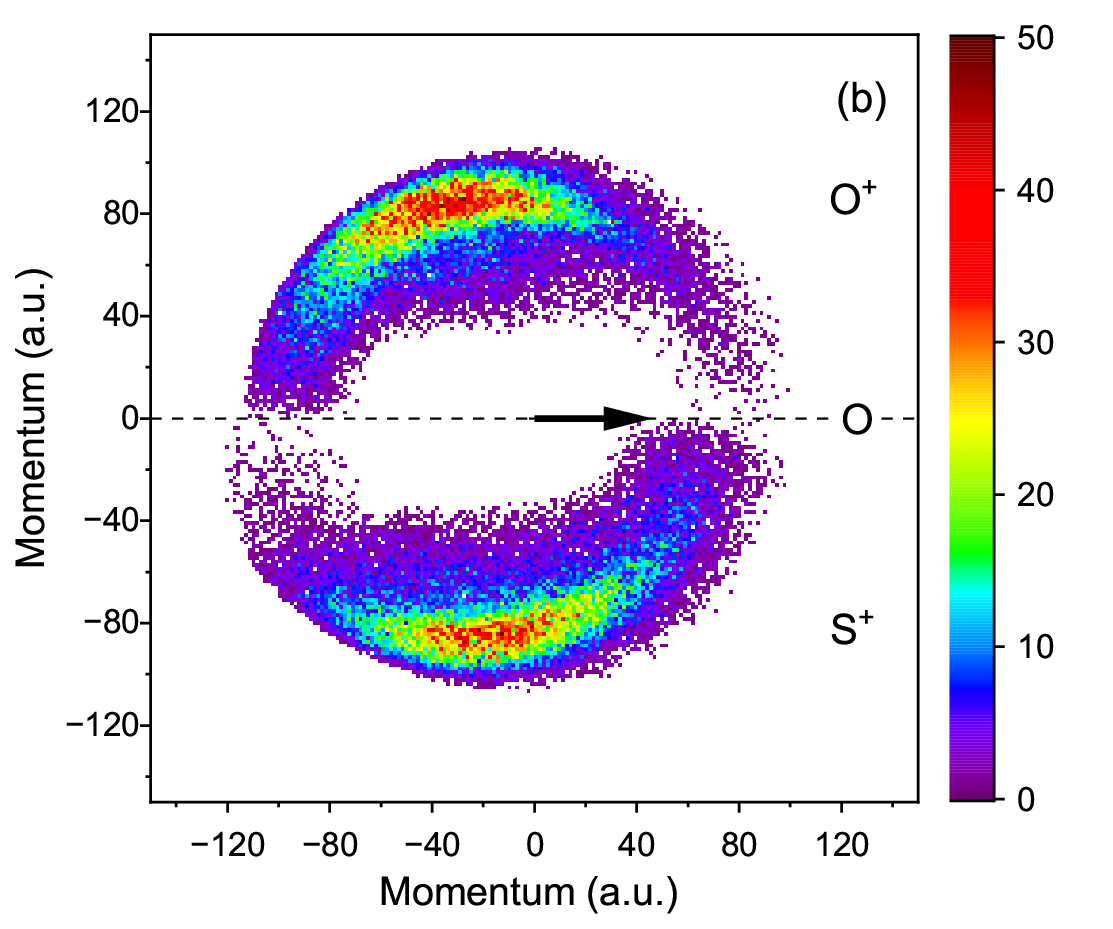}&
  \includegraphics[width=0.325\textwidth]{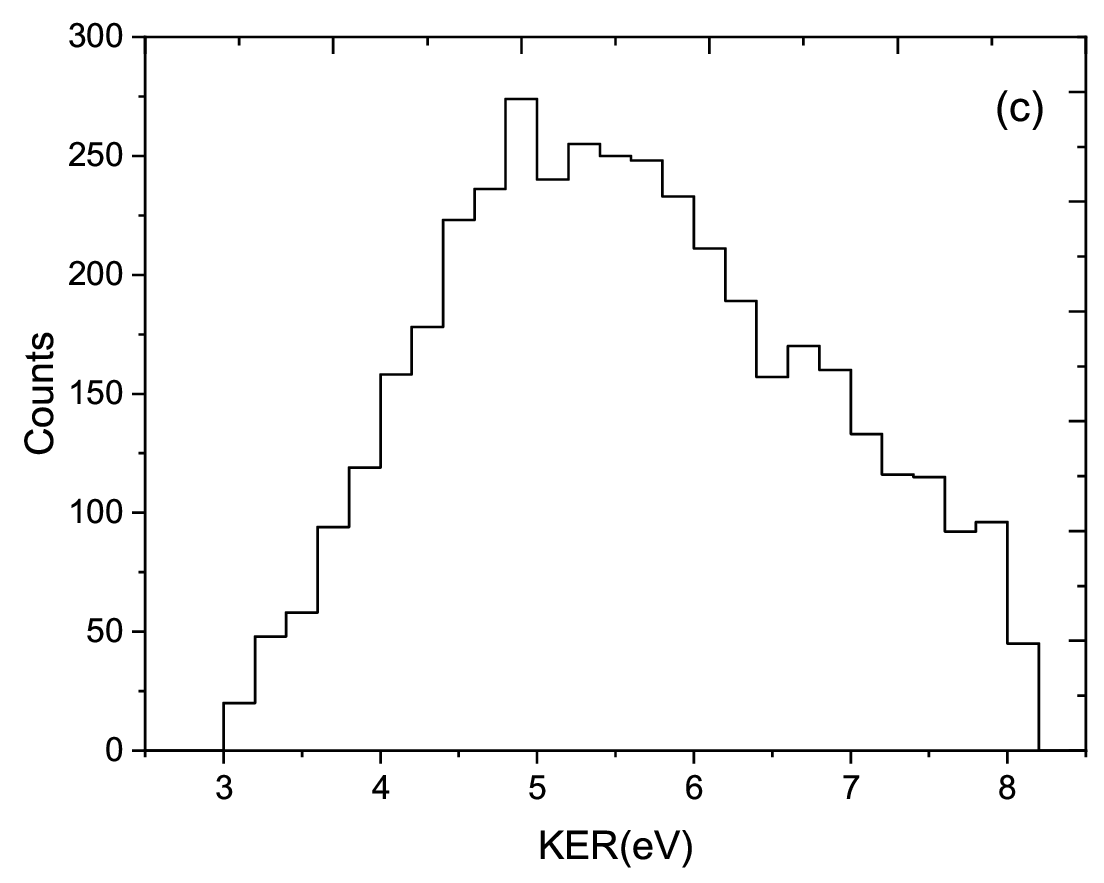} \\
\end{tabular}
     \captionof{figure}{ \justifying (a) 2D plot of angle $\phi$ between $\vec{p}_{O^+S^+}$ and $\vec{p}_O$ versus KER of \ce{O+ + S+} in the center of mass frame of fragment molecular ion \ce{SO^{2+}}. The regions marked by dashed red lines contain distribution of events arising only due to the sequential fragmentation channel \ce{SO2^{2+}  -> SO^{2+} + O -> O+ + S+ + O} (b) Newton diagram for the channel \ce{SO2^{2+}  -> O+ + S+ + O} obtained by applying KER gate of 3 - 8 eV. (c) KER of the second step fragmentation \ce{{OS}^{2+} -> O+ + S+} obtained by applying KER gate of 3 - 8 eV and the gates of 0$^{\circ}$-30$^{\circ}$ and 140$^{\circ}$-180$^{\circ}$ for the angle $\phi$ between $\vec{p}_{O^+S^+}$ and $\vec{p}_O$}\label{fig: O+S+O_R2}
\end{table*}
\begin{table*}
\begin{tabular}{ccc}
  \raisebox{0.065\height}{\includegraphics[width=0.297\textwidth]{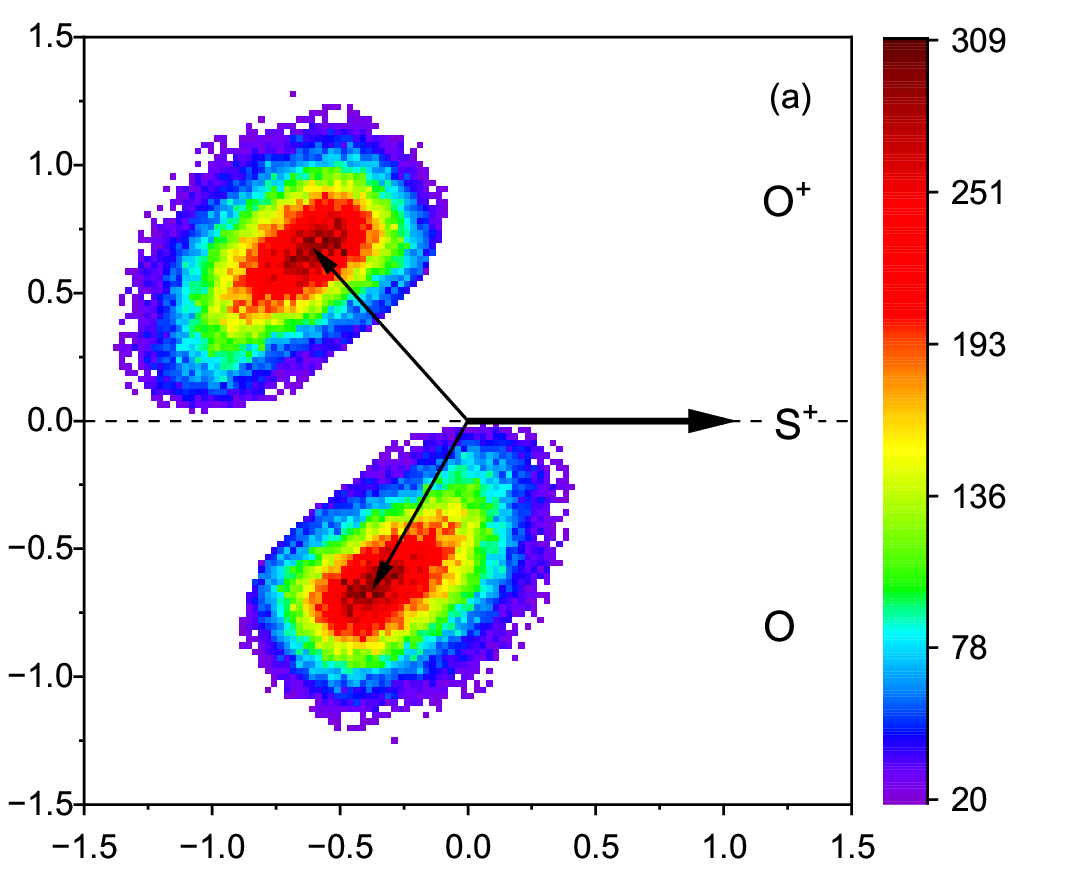}}&
  \includegraphics[width=0.33\textwidth]{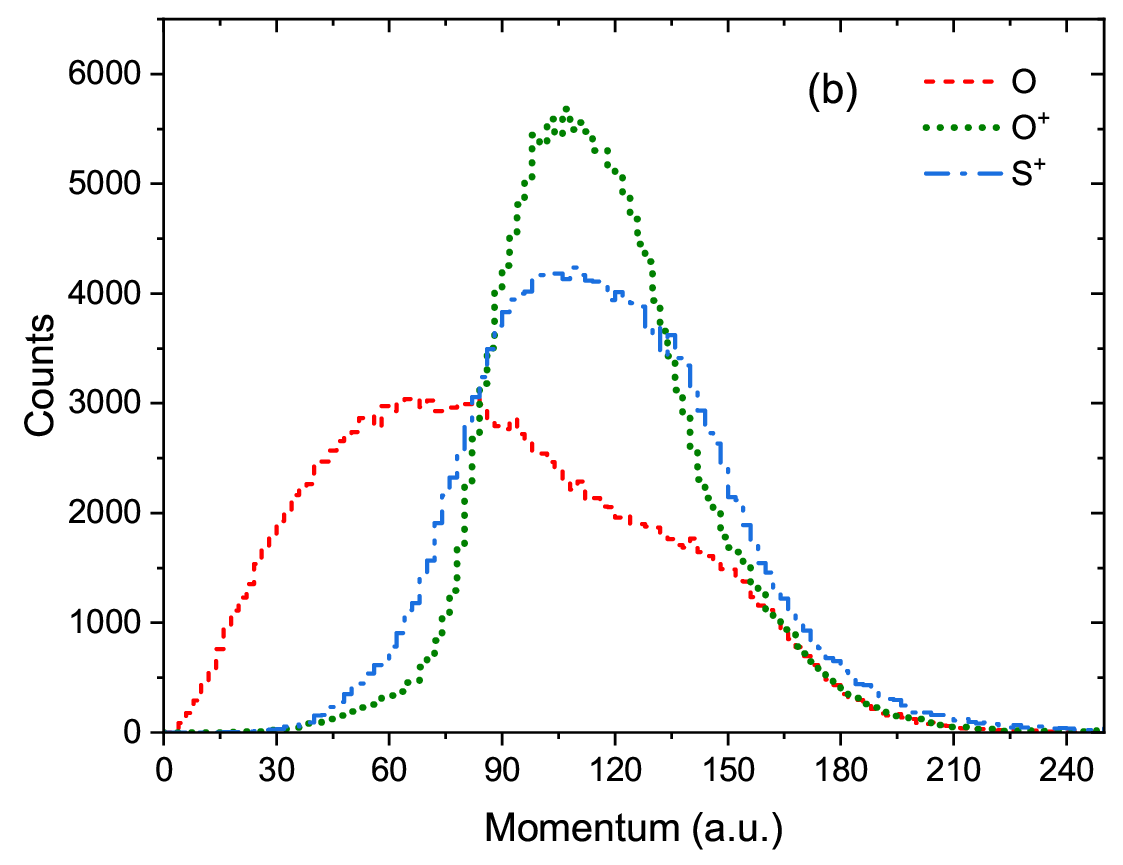} &
  \includegraphics[width=0.34\textwidth]{Figures/fig_5.eps} \\
\end{tabular}
     \captionof{figure}{ \justifying (a) Newton diagram for channel \ce{SO2^{2+}  -> O+ + S+ + O} for full range of KER, (b) Momentum distribution for the channel \ce{SO2^{2+}  -> O+ + S+ + O} and (c) KER distribution for the channel \ce{SO2^{2+}  -> O+ + S+ + O}}\label{fig: O+S+O_R1}
\end{table*} 
\begin{table*}
\begin{tabular}{c}

  \includegraphics[width=0.20\textwidth]{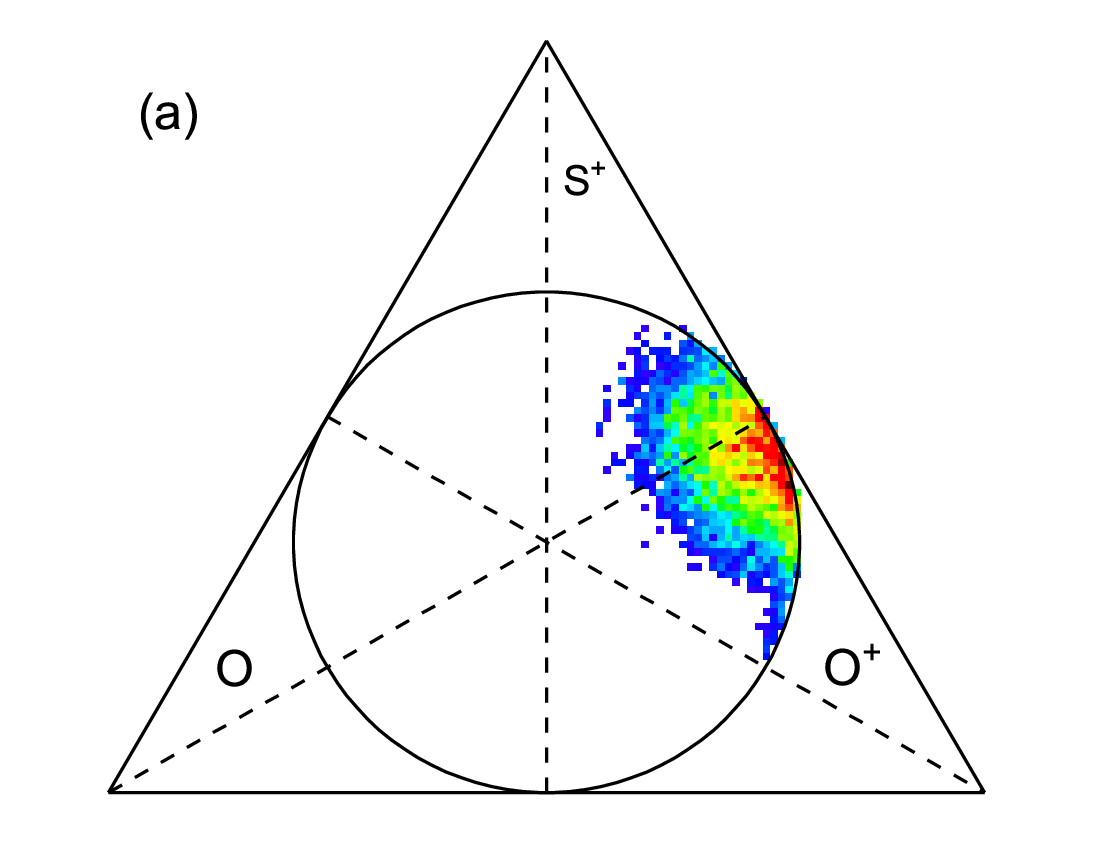}
  \hspace{1.5cm}
  \includegraphics[width=0.15\textwidth]{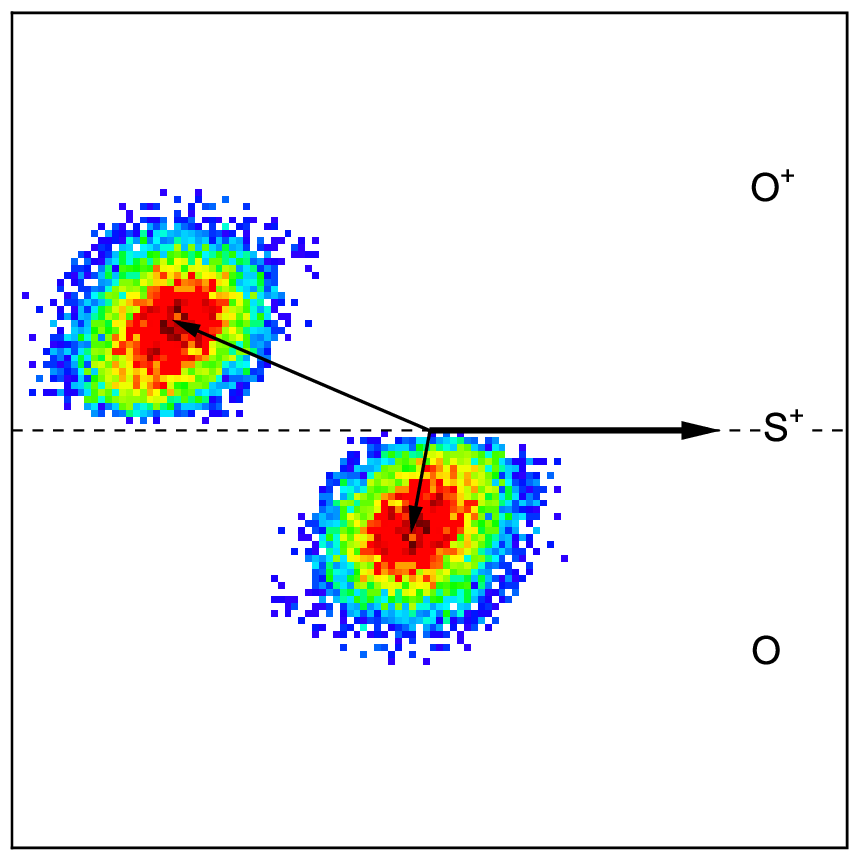}
  \hspace{1.5cm}
  \includegraphics[width=0.205\textwidth]{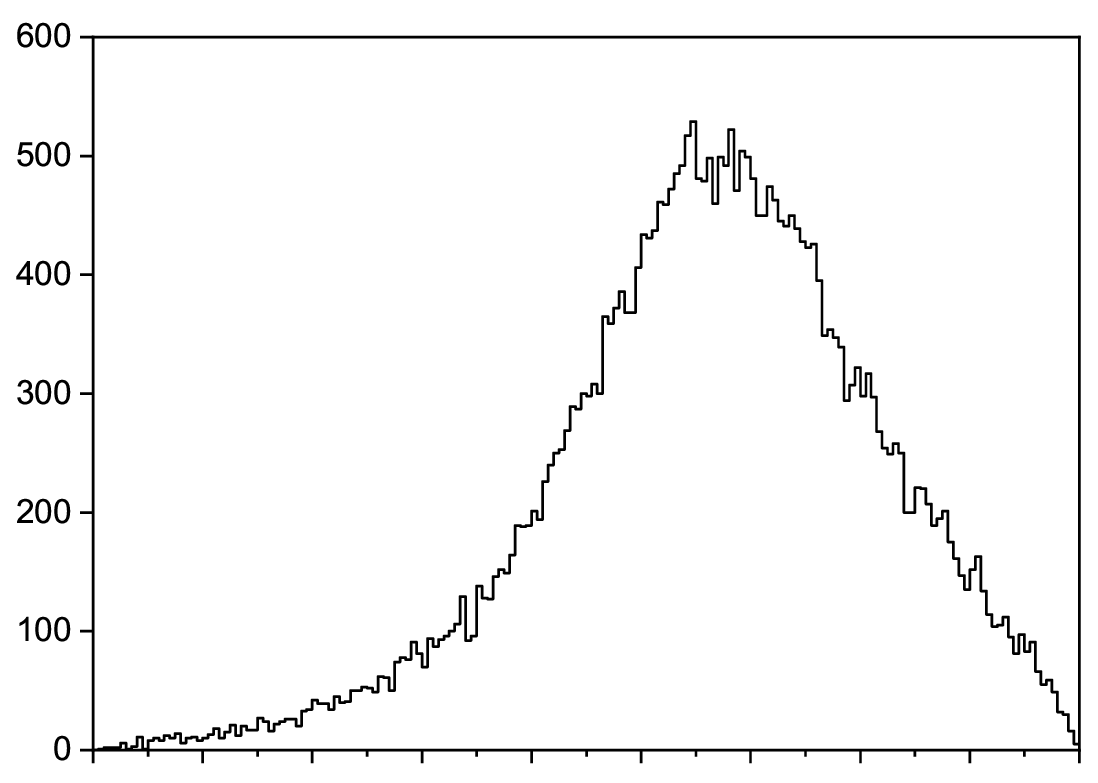}
  \hspace{1.5cm}\\

  \includegraphics[width=0.20\textwidth]{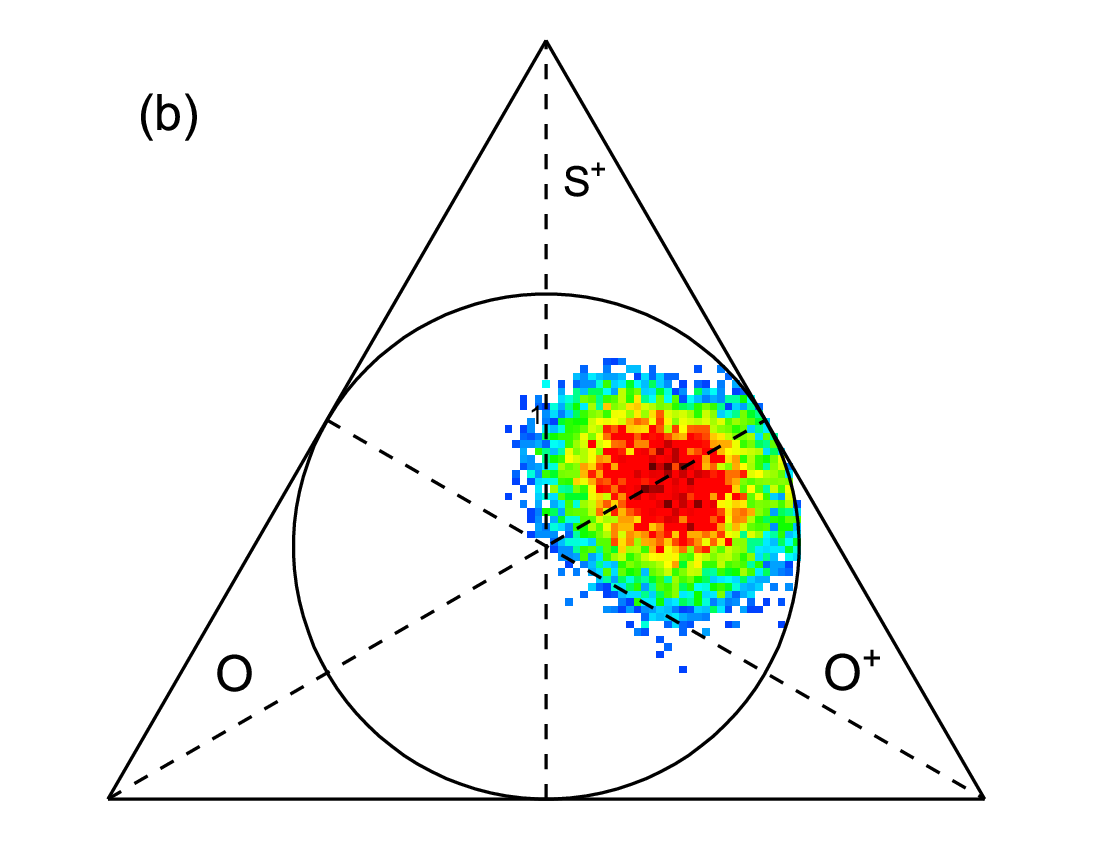}
    \hspace{1.5cm}
  \includegraphics[width=0.15\textwidth]{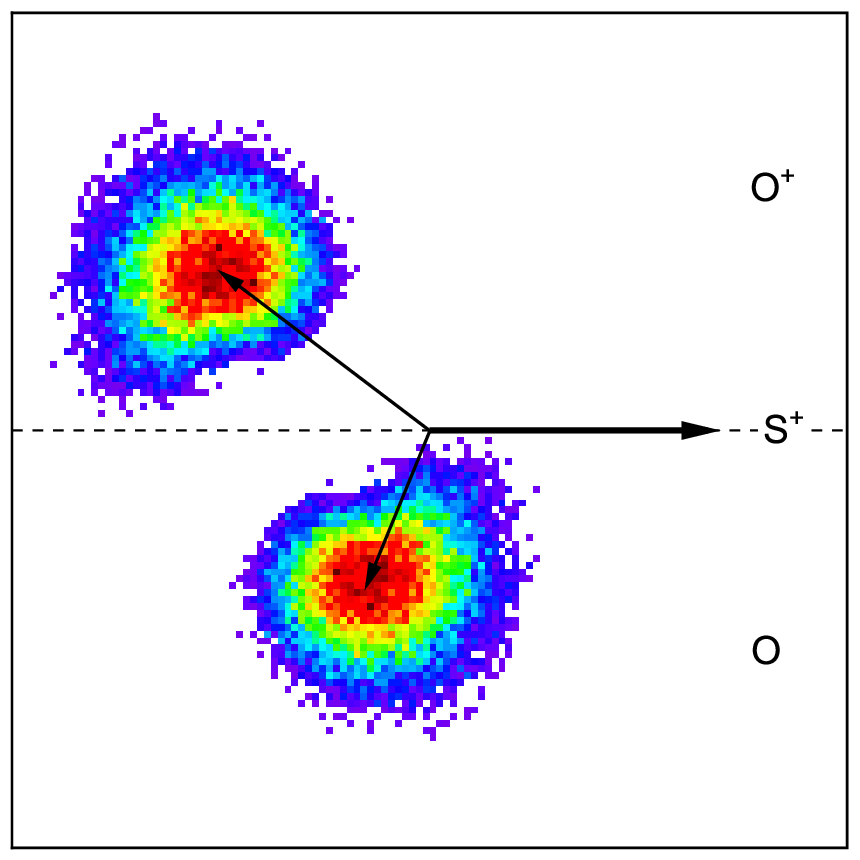}
    \hspace{1.5cm}
  \includegraphics[width=0.205\textwidth]{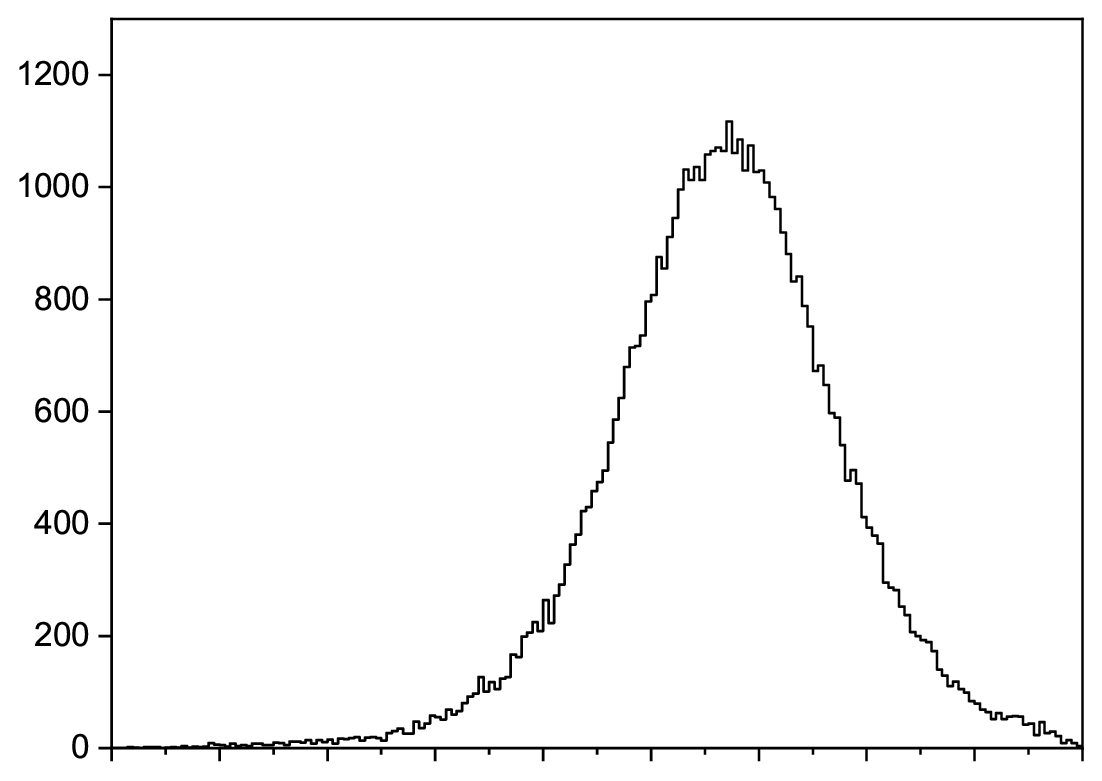}
    \hspace{1.5cm}\\

  \includegraphics[width=0.20\textwidth]{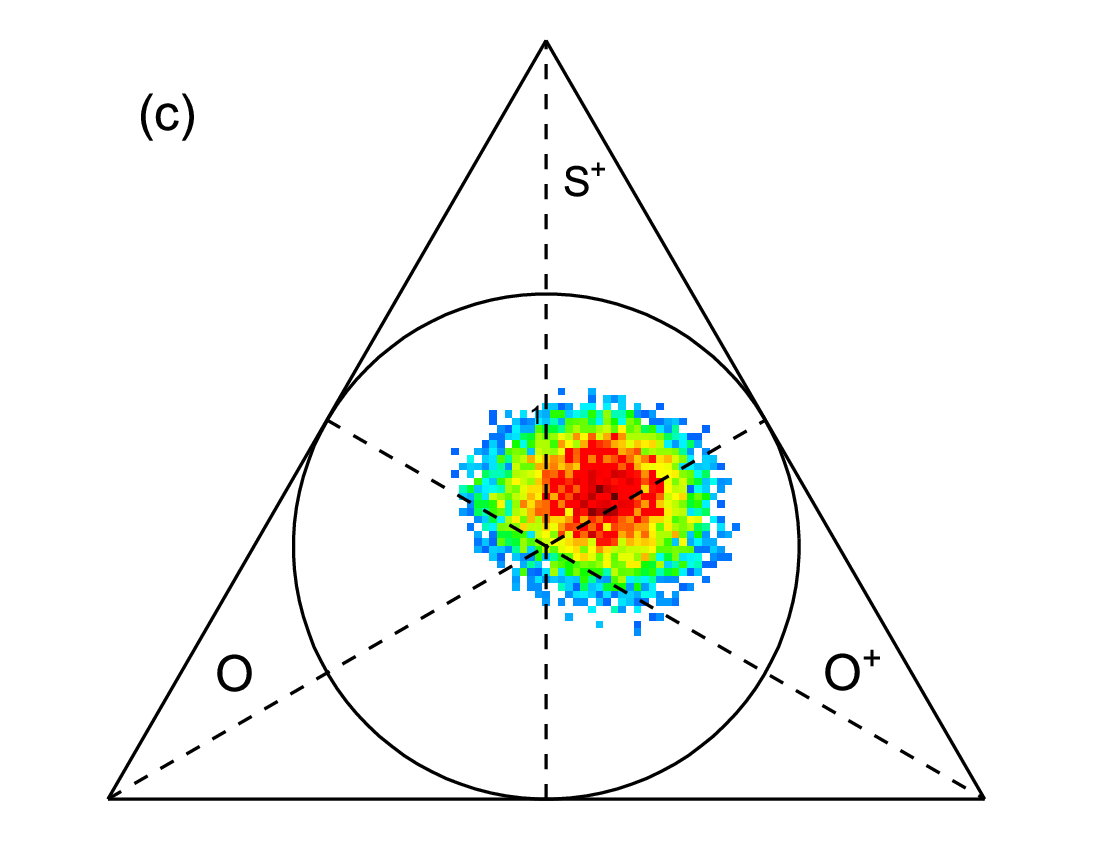}
    \hspace{1.5cm}
  \includegraphics[width=0.15\textwidth]{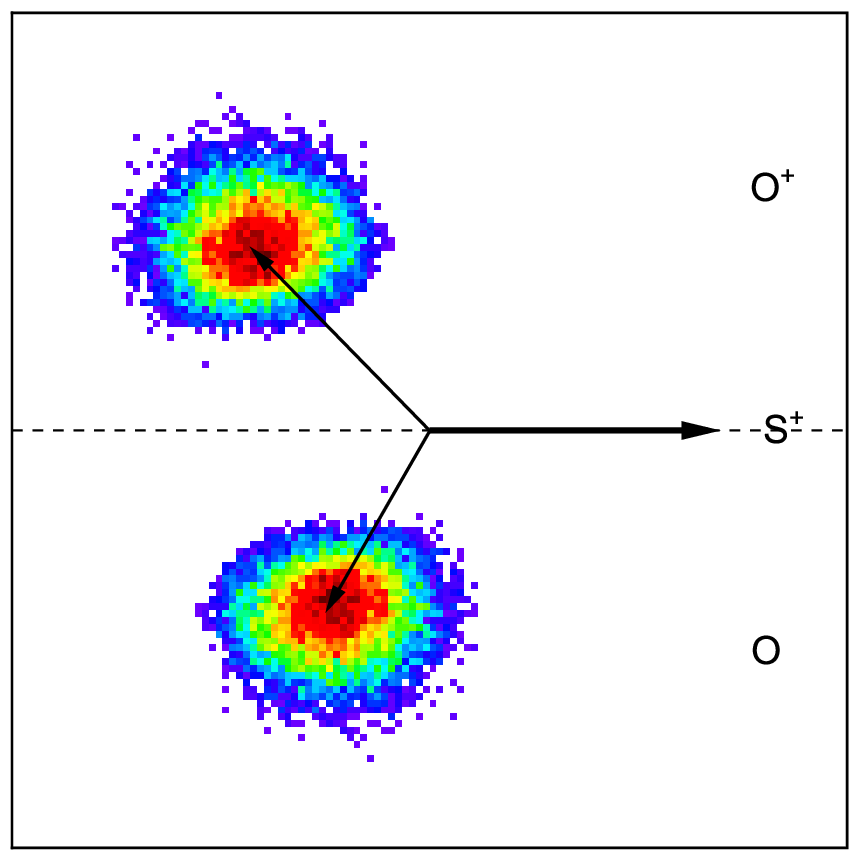}
    \hspace{1.5cm}
  \includegraphics[width=0.205\textwidth]{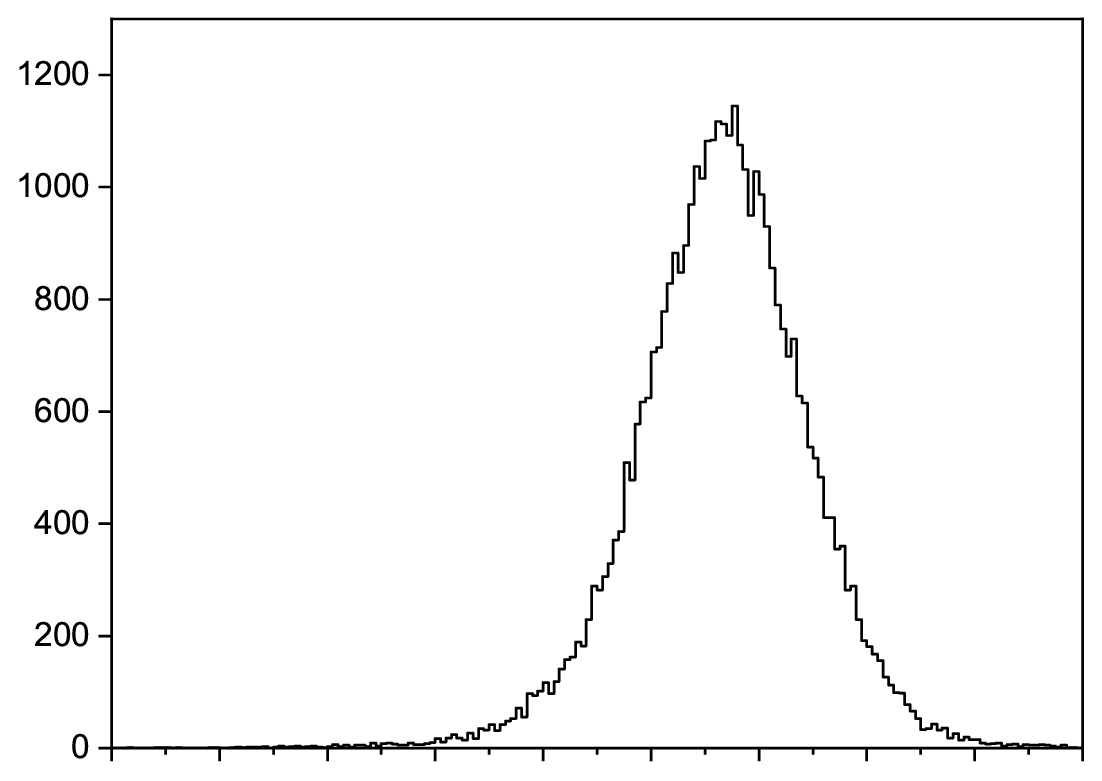}
    \hspace{1.5cm}\\

  \includegraphics[width=0.20\textwidth]{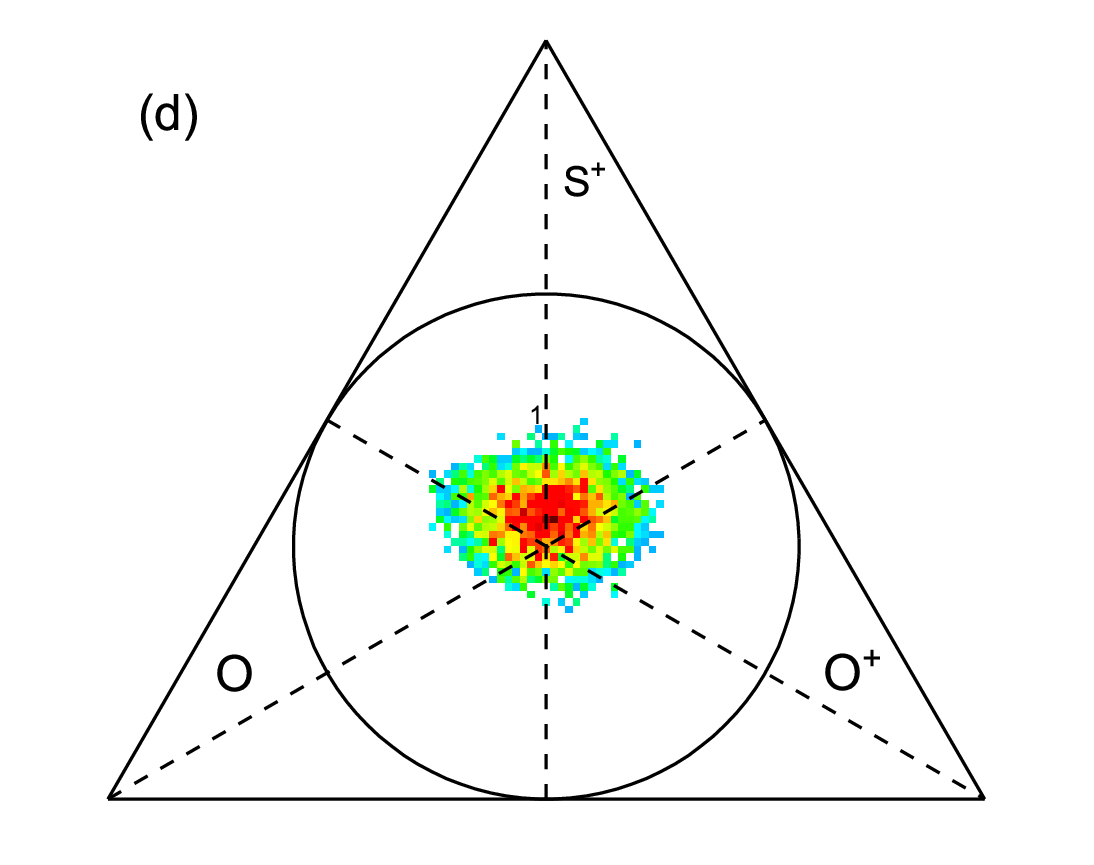}
    \hspace{1.5cm}
  \includegraphics[width=0.15\textwidth]{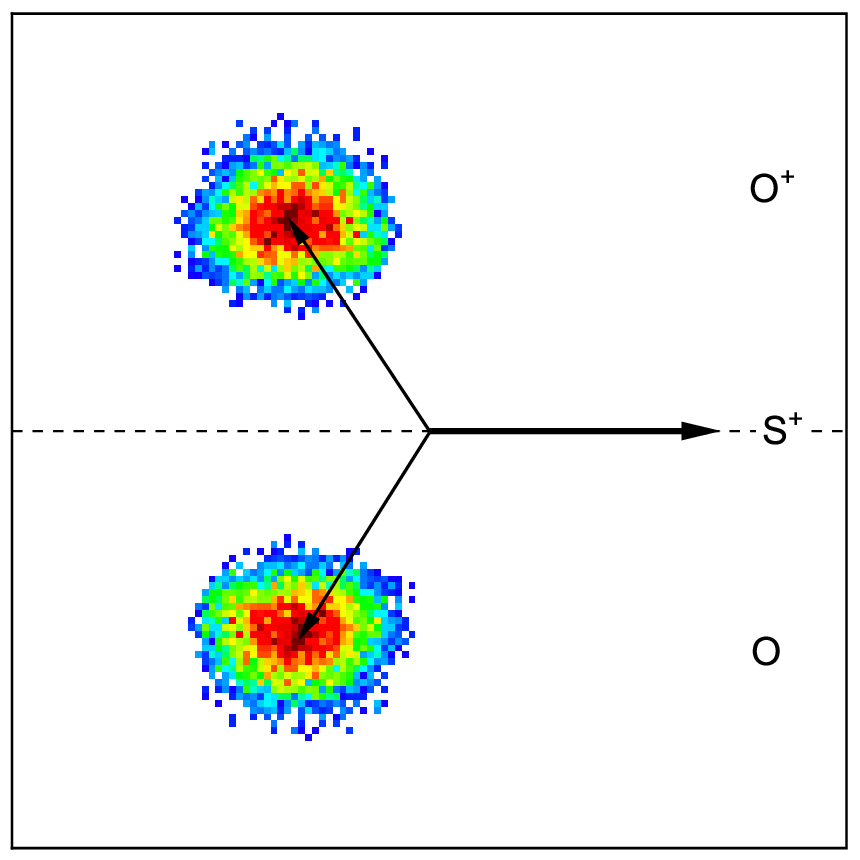}
    \hspace{1.5cm}
  \includegraphics[width=0.205\textwidth]{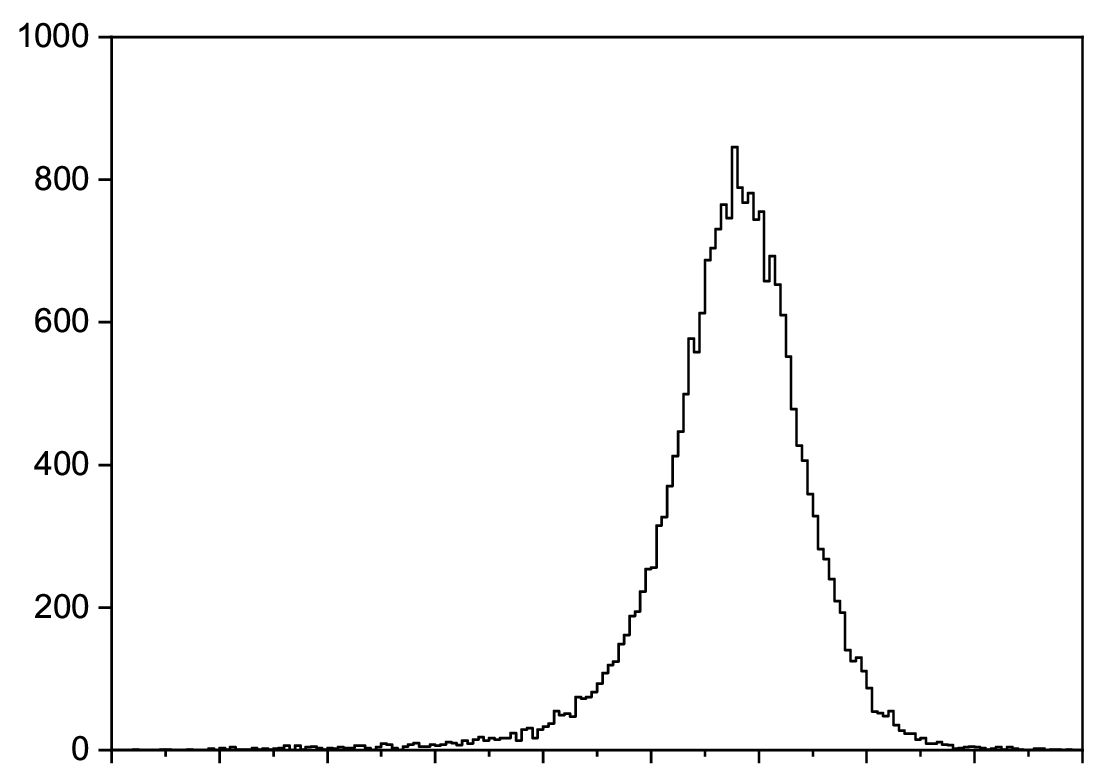}
    \hspace{1.5cm}\\

  \includegraphics[width=0.20\textwidth]{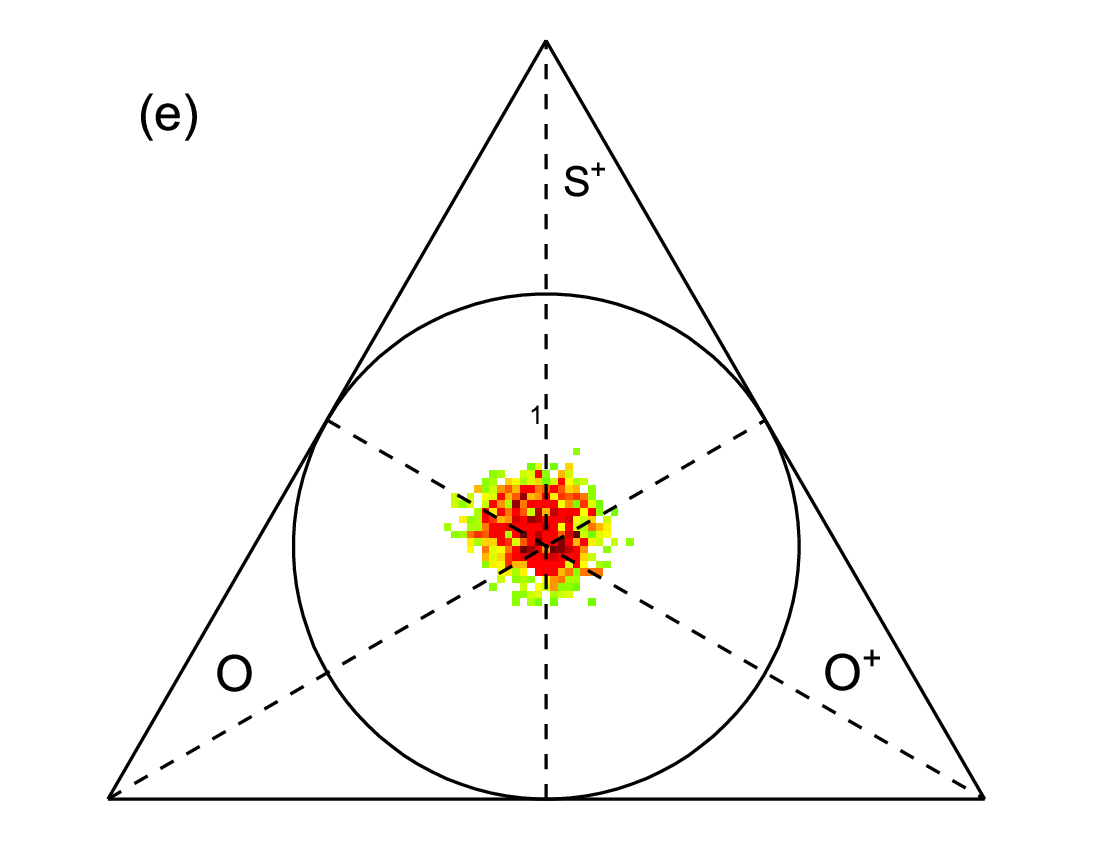}
    \hspace{1.5cm}
  \includegraphics[width=0.15\textwidth]{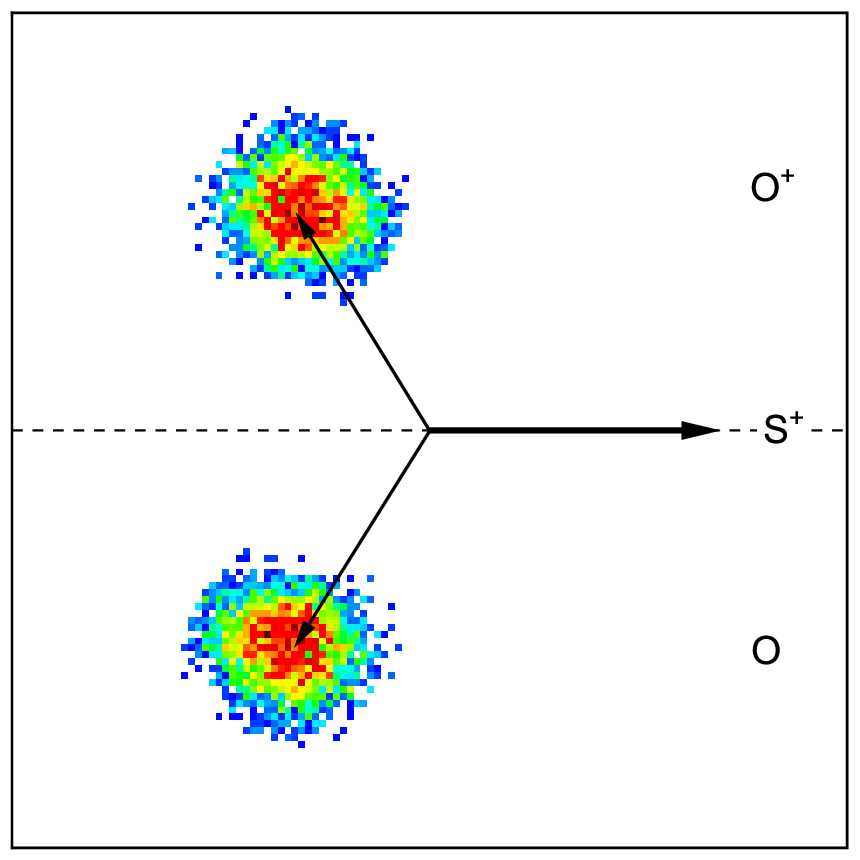}
    \hspace{1.5cm}
  \includegraphics[width=0.205\textwidth]{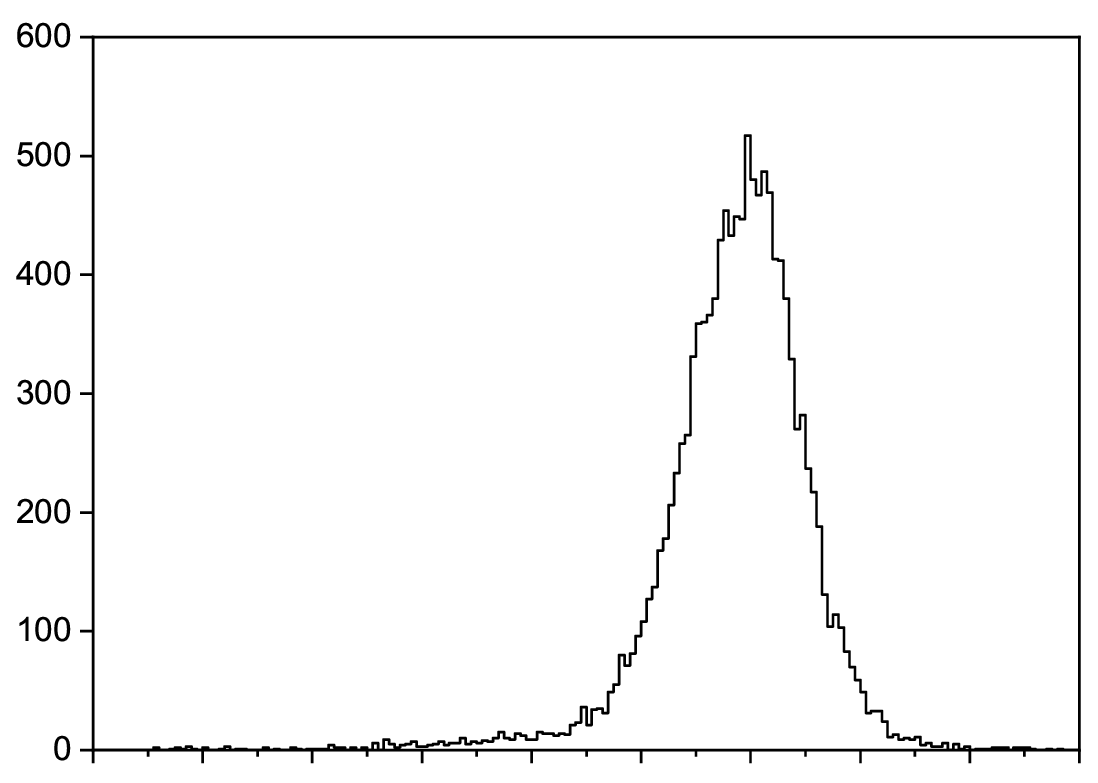}
    \hspace{1.5cm}\\

  \includegraphics[width=0.20\textwidth]{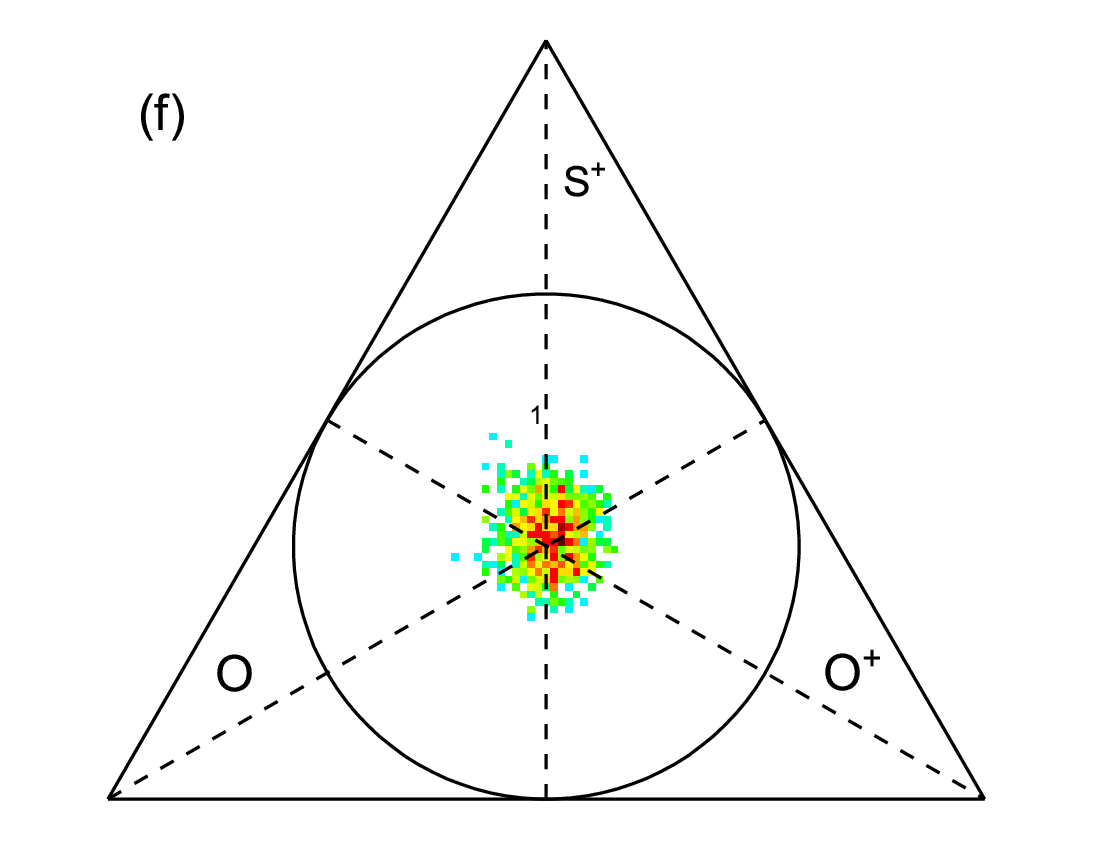}
    \hspace{1.5cm}
  \includegraphics[width=0.15\textwidth]{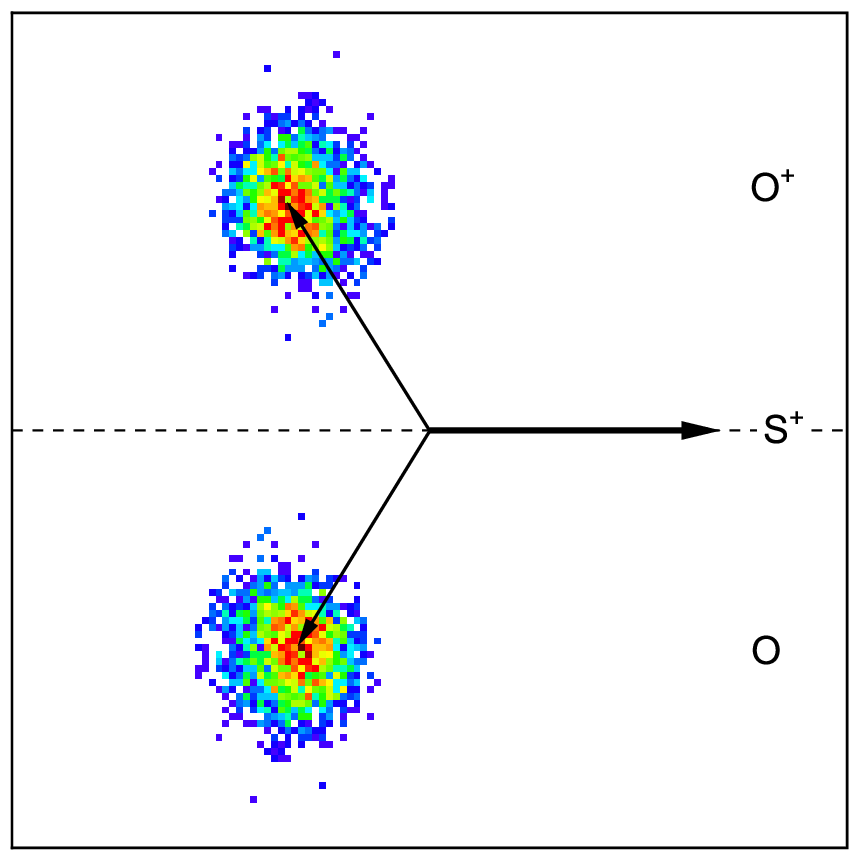}
    \hspace{1.5cm}
  \includegraphics[width=0.205\textwidth]{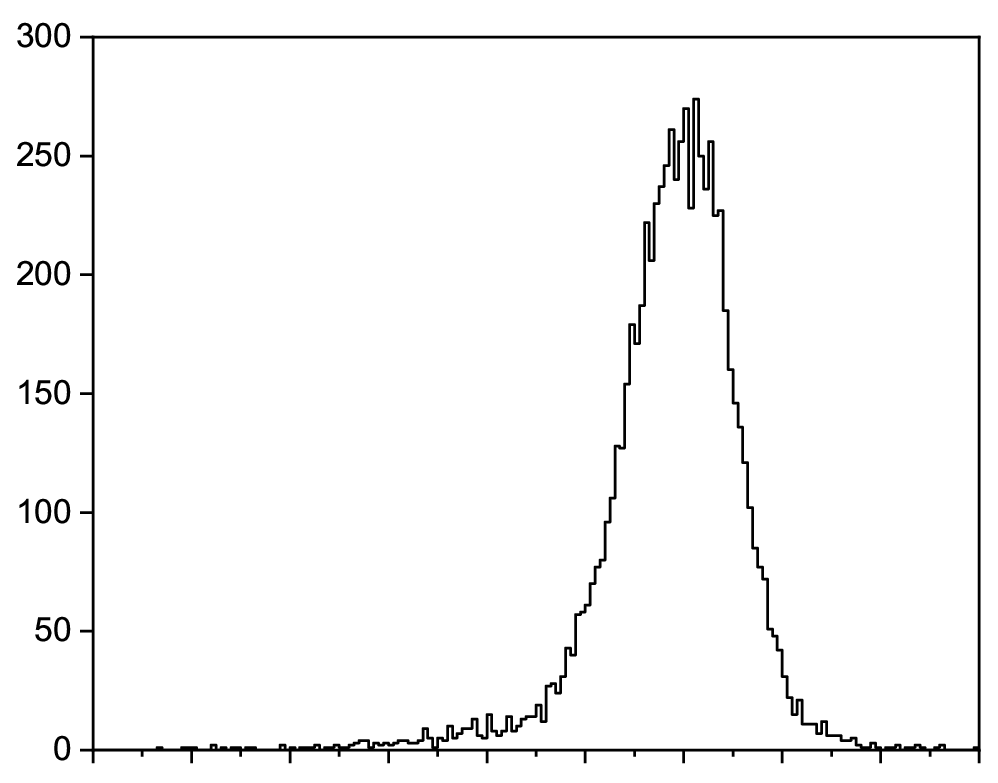}
    \hspace{1.5cm}\\

    \hspace{0.035cm}
  \includegraphics[width=0.20\textwidth]{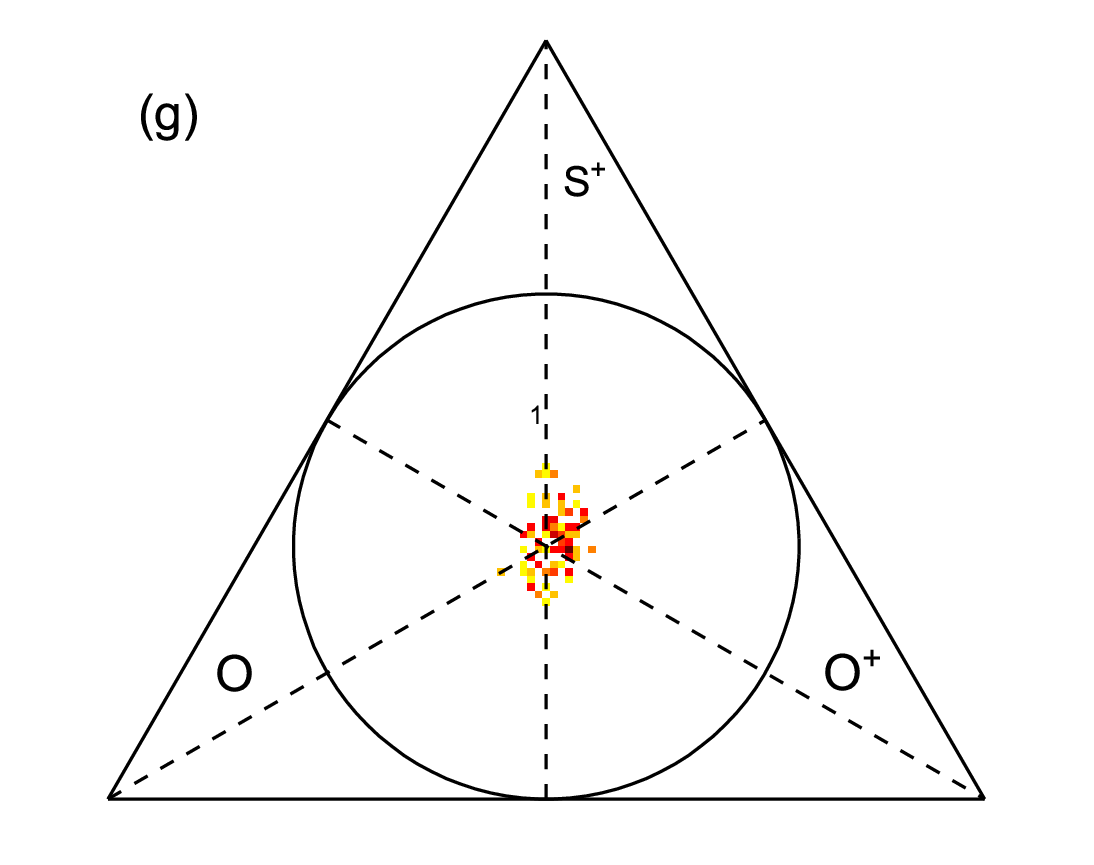}
    \hspace{1.465cm}
  \includegraphics[width=0.15\textwidth]{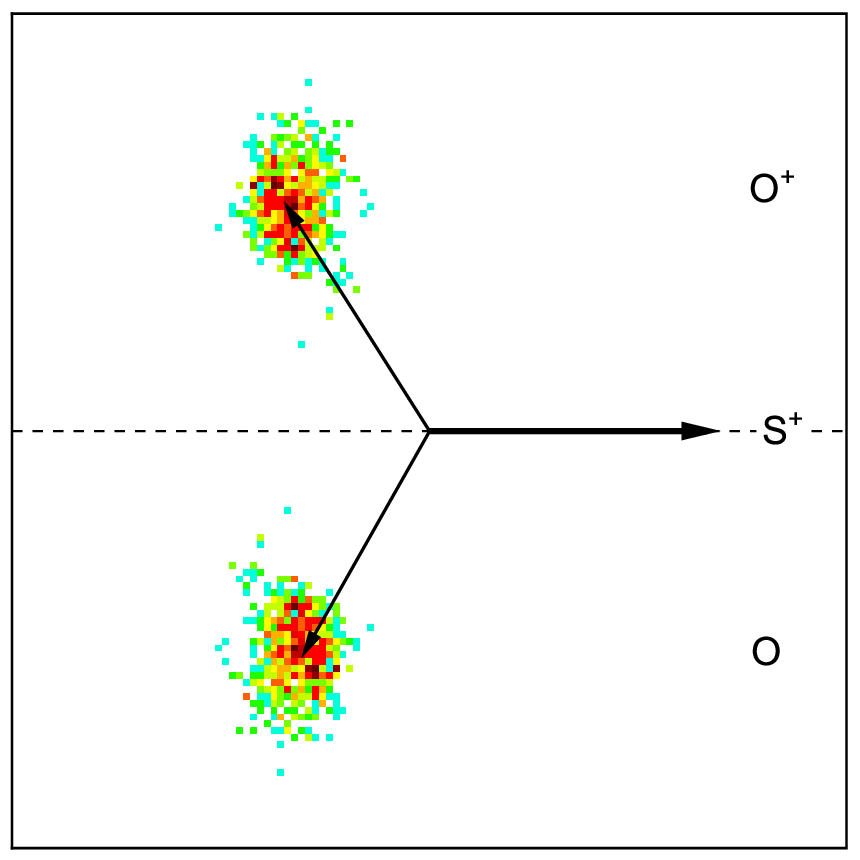}
    \hspace{1.44cm}
   \raisebox{-0.18\height}{\includegraphics[width=0.215\textwidth]{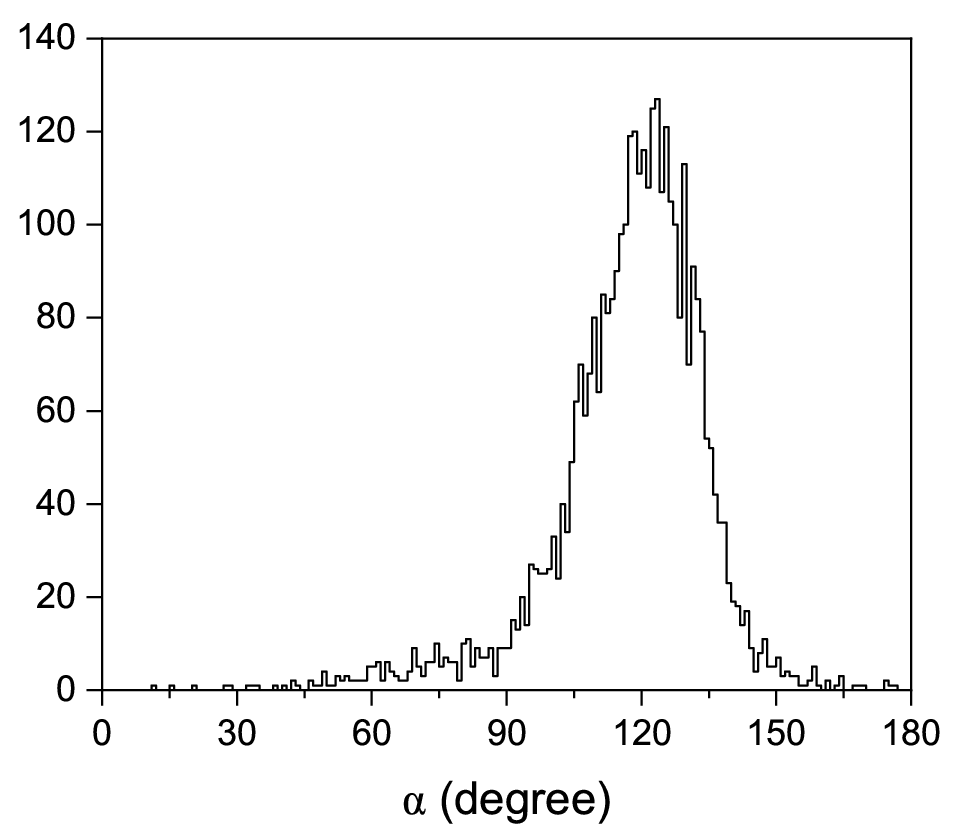}}
    \hspace{-0.5cm}\\

\end{tabular}

    \captionof{figure}{ \justifying Dalitz plot (left panel), the corresponding Newton diagram (middle panel) and the bond angle $\alpha$ for the channel \ce{S+ + O+ + O} for different energy ranges.(a): 0-7.5 eV, (b): 7.5-12.5 eV, (c): 12.5 - 17.5 eV, (d): 17.5 - 22.5 eV, (e): 22.5 - 27.5 eV, (f): 27.5 - 32.5 eV, (g): 32.5 - 40.0 eV.With increasing energy, the configuration approaches that for the concerted fragmentation of triply charged channel \ce{SO2^{3+} -> S+ + O+ + O+}}\label{fig: O+S+O_R1_evolution}
\end{table*}
The figure shows two distinct features, (i) a vertical localization of events corresponding to sequential fragmentation and (ii) a broad triangular distribution of events due to contributions from concerted fragmentation processes. The vertical distribution of events is confined within the KER values of 3 - 8 eV. In fig. \ref{fig: O+S+O_R2}(b) we have plotted the Newton diagram corresponding to this KER range (3 - 8 eV). The Newton diagram shows semicircular structures in the upper and lower quadrants. This is a clear signature of sequential fragmentation. In the first step, the parent \ce{SO2^{2+}} ion dissociates into \ce{SO^{2+}} fragment ion and a neutral \ce{O} atom. The \ce{SO^{2+}} fragment ion further dissociates via Coulomb explosion into \ce{S+} and \ce{O+} ions. This is termed deferred charge separation \cite{eland1987dynamics}. However, the concerted fragmentation pathway also contributes in this kinetic energy range. In order to extract the KER distribution of \ce{SO^{2+}} fragment ion due to sequential fragmentation only, we have selected a narrow range of angles ($\theta$ = 0$^{\circ}$-30$^{\circ}$ and 140$^{\circ}$-180$^{\circ}$) in fig \ref{fig: O+S+O_R2}(a). In this angular range, the contribution from the concerted fragmentation pathway is minimal. The KER distribution for the intermediate \ce{SO^{2+}} ions in their center-of-mass frame can be found using the following equation:
\begin{widetext}
 \begin{equation}
  (KER)^\prime = \frac{1}{2}\left(\ \frac{1}{m_B} + \frac{1}{m_C}\right) \left[ \left(\frac{m_C}{m_{BC}}\right)^2p^2_B + \left(\frac{m_B}{m_{BC}}\right)^2p^2_C
                                 - \frac{2m_B m_C}{m^2_{BC}} \vec{p}_B.\vec{p}_C \right] \label{eqn: NF_KER}
 \end{equation}
 \end{widetext}
 where, the unprimed quantities are in the center-of-mass frame of the parent molecular ion (\ce{SO2^{2+}}). The resulting KER spectrum is shown in fig.~\ref{fig: O+S+O_R2}(c). The most probable kinetic energy value is $\sim$ 5 eV. This is in excellent agreement with the values reported in the literature \cite{eland1987dynamics, uvsor1994, peterson1991spectroscopic, ben2005theoretical}. 
 
 To extract the location of sequential fragmentation events in the Dalitz plot, we simulated the detection of sequential fragmentation events through the momentum spectrometer. The simulations were performed using SIMION 8.0 ion optics package \cite{simion}. The momentum spectrometer was replicated in SIMION \cite{duley2022design}. A code (in C language) was written to generate events corresponding to sequential fragmentation channel. Each event was created using three particles, \ce{S+} ion, \ce{O+} ion, and \ce{O} atom. The \ce{S+} and \ce{O+} ions were generated with equal and opposite momentum values in the center-of-mass frame of \ce{SO^{2+}} dication. The momentum values were distributed around 85 a.u. corresponding to the KER value = 5 eV, as measured in the experiment (see fig. \ref{fig: O+S+O_R2}(c)). The neutral \ce{O} atoms were created with 0.5 eV kinetic energy in the center-of-mass frame of the parent molecule. We generated a total of 650 such events, randomly oriented in space within an interaction region of radius 1 mm. The particles were extracted through the spectrometer and their position and TOF data were recorded at the detector position. The recorded 2D position and TOF values for each event were used to obtain the 3D momenta of fragment ions. These momentum values were further used to generate the corresponding Dalitz plot. The Dalitz plot created using the simulated data for sequential fragmentation shows distribution of events localized parallel to the neutral O axis of the Dalitz plot. This further confirms that \ce{SO2^{2+}} fragmentation into \ce{O+} + \ce{S+} + \ce{O} has contributions from sequential mechanism.

 In fig. \ref{fig: O+S+O_R1}(a) we have shown the Newton diagram for the full range of KER values. 
 The corresponding KER spectrum is shown in fig. \ref{fig: O+S+O_R1}(c). The KER distribution shows two peaks at 8.5 eV and 11.5 eV with a long tail extending up to 40 eV. The peak values are in good agreement with the previously reported values \cite{salen2015complete}. The quantum mechanical states of the parent molecular ion and fragment ions were identified as listed below \cite{dujardin1984double}:
 
 \begin{align}
\ce{ SO2^2+($^{1,3}$B$_2$)  -> S+($^4$S$_u$) + O+($^4$S$_u$) + O($^3$P$_g$)}\label{pathway_1}\\
\ce{ SO2^2+($^{1,3}$B$_2$)  -> S+($^4$S$_u$) + O+($^4$S$_u$) + O($^3$P$_g$)}\label{pathway_2}\\
\ce{ SO2^2+($^{1,3}$B$_1$)  -> S+($^4$S$_u$) + O+($^4$S$_u$) + O($^3$P$_g$)}\label{pathway_3}
\end{align}

To examine the features arising due to concerted fragmentation pathway in more detail, we have selected a range of KER values in steps and constructed the corresponding Dalitz plot and Newton diagram. The bond angle $\alpha$ (see fig. \ref{fig: DMAC}(a)) is also plotted for the respective KER range (see fig. \ref{fig: O+S+O_R1_evolution}). In the KER range 0 - 17.5 eV (fig. \ref{fig: O+S+O_R1_evolution}a - c), the events are distributed towards one side of the Dalitz triangle. This distribution arises due to asymmetric stretch of the bonds. However, the bond angle is close to 120$^\circ$, same as that of the neutral molecule. Therefore, events with low KER values (0 - 17.5 eV) result from asynchronous concerted fragmentation. In the high KER region (17.5  - 40 eV, fig. \ref{fig: O+S+O_R1_evolution}d - g) the events are closer to the center of the Dalitz plot. This implies that the three particles have equal momenta. This is a case of synchronous concerted fragmentation (\ce{SO2^{2+} -> O+ + O+ + S} ) in the geometry of the neutral molecule. We also note that over the entire range of KER, the distribution of the bond angle $\alpha$ peaks around 120$^\circ$. However, this distribution becomes narrower for higher KER values.

A simple Coulomb explosion model predicts KER values of 14.4 eV for two body break-up of \ce{SO^{2+}} fragment ion.  Therefore, KER values above 14 eV cannot be obtained from disintegration of a molecule with a doubly charged core. We also observe that in the higher energy range, the KER distribution matches well with the KER distribution of \ce{SO2^{3+} -> O+ + O+ + S+} reaching the same asymptotic limit ($\sim$ 40 eV) \cite{chen2023fragmentation}. This is attributed to double ionization of \ce{SO2} accompanied by simultaneous excitation of an electron in highly excited state \cite{hsieh1997reaction,ben2005theoretical}. Therefore, the doubly charged molecular ion mimics a triply charged core. As the degree of internal excitation increases, the core becomes triply charged with a distant optical electron. The fragmentation therefore now proceeds in a manner similar to that for synchronous concerted decay of \ce{SO2^{3+}}.

The KER values below 5.5 eV result from a two-step process \cite{hsieh1997reaction}. In the initial step the molecule is singly ionized in an highly excited state. The singly charged molecular ion subsequently autoionizes to a doubly charged molecular ion over a time scale of $\sim$ 100 fs: (\ce{SO2 + H+(1 MeV) -> [O + SO]^{+*} + e^- ->[100 fs] O+ + S+ + O}). This may result in dissociation with lower than expected KER values.

 \section{\label{sec:level7}Conclusion\protect\\ }
Collision of \ce{SO2} with \ce{H^+} results in doubly and triply charged \ce{SO2} molecule, which subsequently fragment due to Coulomb repulsion. Wherever, theoretical calculations are available, the calculated values match well with observations. Of the two possible triple fragmentation channels of \ce{SO2^{2+}} into charged fragments and a neutral, the cross-section of channel \ce{SO2^{2+} -> O+ + O+ + S} is sufficiently large to be analyzed. Excitation of the parent molecular ion results in a change in geometry and symmetry of the molecule which dissociates through different mechanisms. This results in rich structure in the Dalitz plot. These regions were gated out directly from the Dalitz plot to unravel the dissociation pathways. Whenever there was an overlap of different mechanisms, a further gating on the basis of KER was used to separate out the processes clearly. For the reaction \ce{SO2^{2+} -> O+ + S+ + O}, there is considerable overlap of sequential, \ce{SO2^{2+} -> SO^{2+} + O -> O+ + S+ + O}, and concerted fragmentation mechanisms even in the KER distribution. Here, the technique of native frames approach is used to discern the regions and determine the KER in the second step which agrees well with previous observations and calculations.

 \section{\label{sec:level8}Acknowledgment\protect\\ }
S.B.B. is thankful to Dr. Achim Czasch and Dr. Avijit Duley for their help in understanding the CoboldPC software. We also thank Mr. Sahan Sykam for the maintenance of the accelerator facility during the experiments.
 
\nocite{*}

\bibliography{apssamp}

\end{document}